\documentclass[aps,pra,twocolumn,superscriptaddress,a4paper,10pt]{revtex4-2}
\usepackage[T1]{fontenc}
\usepackage[utf8]{inputenc}
\usepackage{array}
\usepackage{multirow}
\usepackage{graphicx}
\usepackage{amsmath,amsfonts,amssymb,color}
\usepackage{amsthm}
\usepackage{mathtools}
\usepackage{leftidx}
\usepackage{xcolor}
\usepackage{dcolumn}
\usepackage{bm}
\usepackage{epstopdf}
\usepackage{epsfig}
\usepackage{environ}
\usepackage{pdfcomment}
\usepackage{float}
\usepackage{tabularray}
\usepackage{hyperref}

\usepackage{esint}
\usepackage{physics}
\usepackage[english]{babel}

\hypersetup{colorlinks=true,citecolor=blue,
	linkcolor=blue,urlcolor=blue,pdfstartview=FitH,
	bookmarksopen=true}
\usepackage{cleveref}

\begin{document}
	
	\title{Probing non-ergodicity and symmetry via coherent forward scattering in a shaken rotor}

    \author{F. Arrouas}
\affiliation{Laboratoire Collisions Agrégats Réactivité, Université de Toulouse, CNRS, 31062 Toulouse, France}

    \author{J. H\'ebraud}
\affiliation{Laboratoire de Physique Théorique, Université de Toulouse, CNRS, UPS, France}

    \author{N. Ombredane}
\affiliation{Laboratoire Collisions Agrégats Réactivité, Université de Toulouse, CNRS, 31062 Toulouse, France}

    \author{E. Flament}
\affiliation{Laboratoire Collisions Agrégats Réactivité, Université de Toulouse, CNRS, 31062 Toulouse, France}

    \author{D. Ronco}
\affiliation{Laboratoire Collisions Agrégats Réactivité, Université de Toulouse, CNRS, 31062 Toulouse, France}

    \author{N. Dupont}
\affiliation{Center for Nonlinear Phenomena and Complex Systems, Université Libre de Bruxelles, CP 231, Campus Plaine, 1050 Brussels, Belgium}
\affiliation{International Solvay Institutes, 1050 Brussels, Belgium}

    \author{G. Lemarié}
\affiliation{MajuLab, CNRS-UCA-SU-NUS-NTU International Joint Research Unit, Singapore}
\affiliation{Centre for Quantum Technologies, National University of Singapore, Singapore}
\affiliation{Department of Physics, National University of Singapore, Singapore 117542, Singapore}
\affiliation{Laboratoire de Physique Théorique, Université de Toulouse, CNRS, UPS, France}
    \author{B. Georgeot}
\affiliation{Laboratoire de Physique Théorique, Université de Toulouse, CNRS, UPS, France}
    \author{Ch. Miniatura}
\affiliation{Institut de Physique de Nice, Université Côte d’Azur, CNRS, 06200 Nice, France}
\affiliation{Centre for Quantum Technologies, National University of Singapore, Singapore}
    \author{J. Billy}
\affiliation{Laboratoire Collisions Agrégats Réactivité, Université de Toulouse, CNRS, 31062 Toulouse, France}

    \author{B. Peaudecerf}\email[Corresponding author: ]{bruno.peaudecerf@cnrs.fr}

\affiliation{Laboratoire Collisions Agrégats Réactivité, Université de Toulouse, CNRS, 31062 Toulouse, France}

    \author{D. Guéry-Odelin}
\affiliation{Laboratoire Collisions Agrégats Réactivité, Université de Toulouse, CNRS, 31062 Toulouse, France}
    
    \date{\today}

\begin{abstract}

The Coherent Backscattering (CBS) peak is a well-known interferential signature of weak localization in disordered or chaotic systems.
More recently, a second interference feature---the Coherent Forward Scattering (CFS) peak---was predicted to emerge in the regime of strong localization. However, it has never been directly observed.
Here we report the first direct observation of the CFS peak and demonstrate its dual role as a signature of non-ergodicity and as a probe of symmetries in quantum chaotic systems. Using a shaken rotor model realized with a Bose-Einstein condensate (BEC) of ultracold atoms in a modulated optical lattice, we investigate dynamical localization in momentum space. The CFS peak emerges in the position distribution as a consequence of non-ergodic dynamics, while its growth timescale reveals the underlying localization length. By finely tuning the modulation, we control time-reversal and parity symmetries and measure their distinct effects on both CBS and CFS peaks. Our results highlight the strong link of both the  temporal growth and contrast of the CFS with symmetry and localization, making it a robust probe of these properties. This work opens new directions for characterizing non-ergodicity and symmetries in quantum chaotic or disordered systems, with possible applications in many-body localization and chaos.

\end{abstract} 
 
	\maketitle

\begin{figure}[h]
    \centering
    \includegraphics[width=0.45\textwidth]{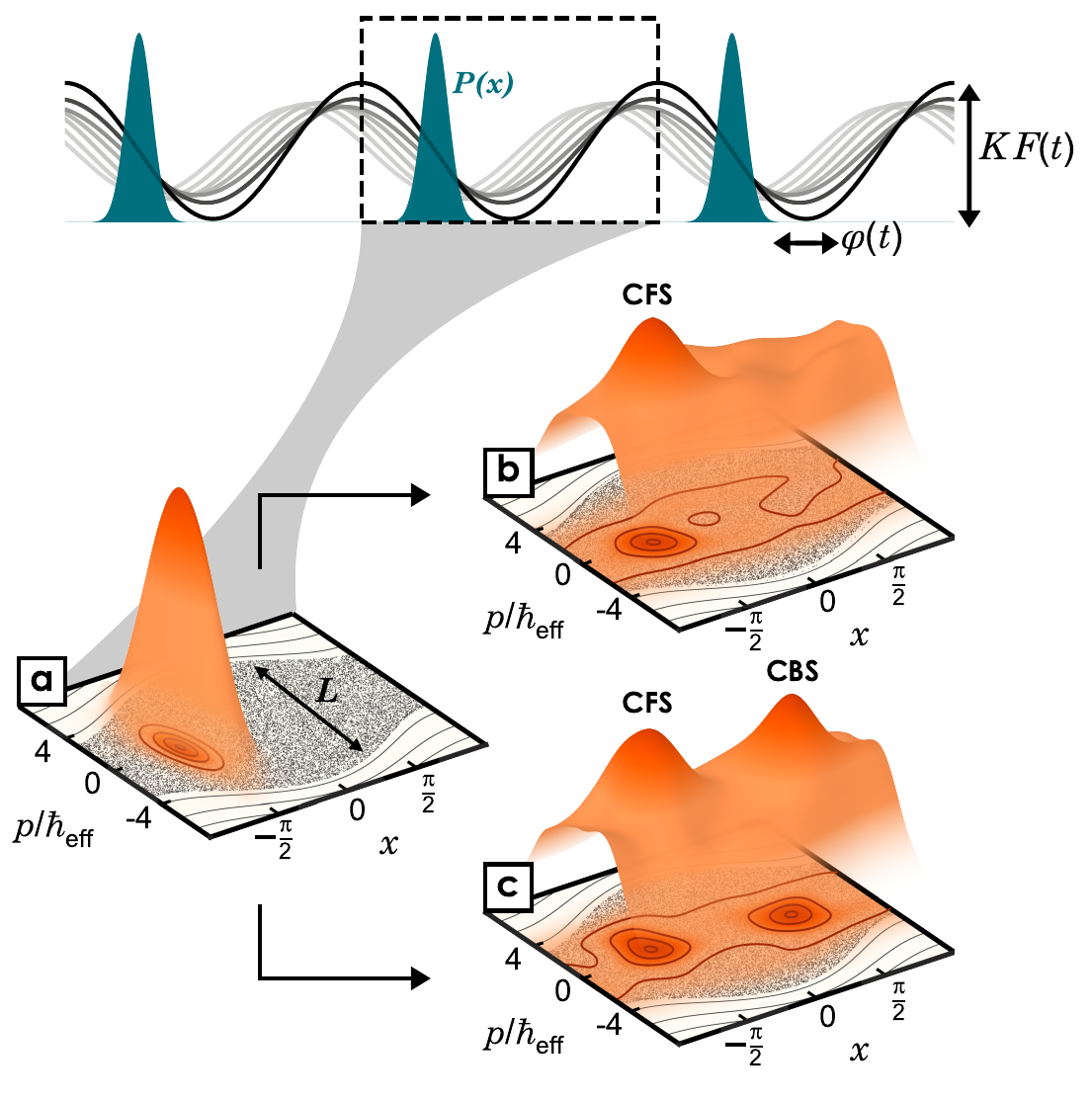}
    \caption{Top: Sketch of the initial atomic density in a 1D lattice, which is subsequently periodically modulated in position ($\varphi(t)$) and amplitude ($F(t)$). 
    Bottom: Experimentally measured Husimi quasi-distributions of (\textbf{a}) the initial squeezed Gaussian state peaked at $x=-\pi/2$,  shown above the classical stroboscopic phase portrait, which highlights the classically chaotic region of size $L_p=L\hbar_\mathrm{eff}$. After undergoing periodic modulation for an evolution time $t>t_H$ sufficient for localization to set in, the coherent scattering peaks are measured: (\textbf{b}) In the absence of an appropriate time-reversal symmetry, only the CFS peak is observed. (\textbf{c}) When the dynamics possess the appropriate symmetry, both the CFS and the CBS peak, at $x=\pm\pi/2$ are observed. The Husimi distributions are obtained from a tomographic full quantum state reconstruction, combined for (\textbf{b},\textbf{c}) with an average over realizations of chaotic dynamics (see text and Methods for details). Parameters are ($\mathbf{b}$) $\hbar_{\rm eff}=9.5$, $K=110\pm7$, $L=8.76\pm0.38$, $\xi_{\rm loc}=31.9\pm9.6$, $M=10$, ($\mathbf{c}$) $\hbar_{\rm eff}=9.55$, $K=113\pm7$, $L=8.99\pm0.20$, $\xi_{\rm loc}=38.9\pm 17.6$, $M=10$.}
    \label{fig:Fig1_schematicillustration}
\end{figure}

Dynamical chaos~\cite{ott2002chaos} plays a central role in understanding how physical systems governed by deterministic, time-reversal-invariant equations can give rise to probabilistic descriptions, which lie at the heart of statistical physics~\cite{gallavottiStatisticalMechanics1999}. Chaotic dynamics generally imply ergodicity, a regime in which the time spent in each accessible phase space region is proportional to its volume. Recently, chaotic dynamics and its associated relaxation towards an equilibrium distribution have attracted renewed attention in closed quantum systems, such as ultracold atomic gases, trapped ions, and spin qubits~\cite{d2016quantum, Gogolin_2016}. These systems, remarkably well-isolated from their environment, provide ideal platforms for exploring the fundamental question of whether a quantum system can relax to a stationary equilibrium solely through unitary dynamics~\cite{ueda2020quantum}.

Consequently, growing interest has focused on mechanisms that allow evading this relaxation, \emph{i.e.,} non-ergodicity. This can occur classically through mixed dynamics, where regular trajectories lie within a chaotic sea in phase space, or due to classical barriers or inhomogeneous chaotic properties~\cite{ott2002chaos, bohigas1993manifestations}. In the quantum realm, Anderson localization, which arises from the interplay between disorder or chaotic diffusion and interference effects~\cite{PhysRev.109.1492, RevModPhys.80.1355, anderson201050, santhanam2022quantum}, and its recent generalization, many-body localization~\cite{RevModPhys.91.021001, alet2018many, tikhonov2021anderson, sierant2025many}, have been identified as key mechanisms for non-ergodicity. Other mechanisms, particularly in many-body quantum systems, have also been recently highlighted, e.g., quantum many-body scars~\cite{serbyn2021quantum}, fragmentation of Hilbert space~\cite{zhaoObservationQuantumThermalization2024, adlerObservationHilbertSpace2024}, dynamical symmetries~\cite{bucaNJP2020}, and prethermalization in periodically driven systems~\cite{hoQuantumClassicalFloquet2023}. Many-body localization also plays a role in stabilizing topological features such as edge states~\cite{RevModPhys.82.3045} - which are themselves a form of non-ergodicity - at finite temperature~\cite{parameswaran2018many}.

In ergodic and non-ergodic systems alike, dynamics is affected by symmetries. Even in systems exhibiting fully chaotic classical dynamics, time-reversal symmetry can significantly affect quantum transport. A striking example is the disappearance of the coherent backscattering (CBS) peak without time-reversal symmetry~\cite{akkermans2007mesoscopic, hainaut2018controlling, CBSAtom1}. Likewise, the spectral statistics of quantum chaotic or disordered systems crucially depend on this symmetry~\cite{RevModPhys.80.1355, haake1991quantum, bohigasCharacterizationChaoticQuantum1984}. 

A central challenge is thus to identify clear, unambiguous signatures that distinguish ergodic from non-ergodic behavior, as well as a system's symmetry properties. While imbalance or entanglement entropy growth have been highlighted in the many-body regime~\cite{RevModPhys.91.021001, serbyn2021quantum}, another signature of non-ergodicity was recently discovered in the non-interacting Anderson localization context: the Coherent Forward Scattering (CFS) peak~\cite{PhysRevLett.109.190601, PhysRevA.90.063602, PhysRevA.90.043605, PhysRevA.95.043626, PhysRevA.95.041602, Martinez_2023, PhysRevResearch.6.L012021,MicklitzPRL2014}. This peak arises from interference effects immune to disorder averaging, when quantum dynamics becomes effectively confined in phase space. It is therefore associated with strong non-ergodicity, which may be induced by classical dynamical barriers or by interference-induced strong localization effects.  

In spatially localized systems, the CFS peak emerges in the forward direction of the final momentum distribution of a plane wave initially launched into a disordered medium~\cite{PhysRevA.90.063602}. More generally, it arises at the initial coordinate in the reciprocal space of the one where localization occurs~\cite{PhysRevA.95.043626}. In the context of Anderson localization~\cite{PhysRevLett.109.190601}, the CFS peak only appears after localization has set in and is absent in the diffusive transport regime~\cite{PhysRevA.95.041602}. Remarkably, it can also detect highly non trivial forms of non-ergodic behavior, such as quantum multifractality, which arises at the Anderson transition between localized and delocalized phases~\cite{PhysRevA.95.041602,Martinez_2023}. Finally, it crucially depends on the symmetries of the system, either through its contrast or its growth dynamics~\cite{PhysRevResearch.6.L012021}, making it a prime probe for non-ergodicity and symmetries.

Cold-atom systems, with their high degree of experimental control and isolation, constitute a versatile platform for exploring non-ergodic dynamics.  They have proven particularly fruitful for investigating localization phenomena, from 1D Anderson localization in a disordered potential~\cite{Billy08} to the measurement of the Anderson transition in a kicked-rotor atomic system~\cite{Chabe08,Madani24}. They also enabled the observation of localization-related phenomena such as the enhanced return to the origin (ERO)~\cite{hainaut2017ERO} and the boomerang effect~\cite{Sajjad22}. In ERO measurements in a quasi-periodic kicked rotor~\cite{hainaut2018controlling}, separate contributions to the ERO from CBS- and CFS-related interference effects could be separated from a background, through a careful choice of modulation providing an effective synthetic time dimension. While they allow to measure the growth of these interference effects, the ERO signatures~\cite{Prigodin1994, Weaver2000} are distinct from the coherent scattering peaks, lacking access, for example, to the peak width, which carries information on localization - and no direct measurement of a genuine CFS peak has been achieved to date.

In this Letter, we report the first direct measurement of the CFS peak in a cold-atom system and show that it is not only a hallmark of non-ergodicity but can also probe the underlying symmetries via its asymptotic contrast. We implement a shaken rotor model, using a Bose-Einstein condensate (BEC) in a modulated optical lattice. Like the kicked rotor, it exhibits chaotic classical dynamics and, in the quantum regime, dynamical localization in momentum space~\cite{mooreObservationDynamicalLocalization1994e}, giving rise to coherent scattering peaks in the position distribution. Crucially, the shaken rotor allows us to tailor the symmetries of the dynamics, while averaging over realizations of classical chaos. Leveraging the high degree of control in the lattice system, we prepare an initially narrow position distribution that subsequently undergoes chaotic dynamics in the modulated lattice, giving rise to the CFS. We perform a full state reconstruction that allows to visualize the scattering peaks over phase space, and characterize their width. In addition to controlling symmetries, the modulation parameters also determine the localization scale, enabling us to tune between localization-dominated and classically-bounded regimes. We measure the symmetry-dependent suppression or enhancement of the CFS and CBS scattering peaks and highlight the relation between the localization length, symmetries, and the growth timescale of the CFS.

\begin{figure*}[t]
    \centering
    \makebox[\textwidth][c]{%
    \includegraphics[width=1.\textwidth]{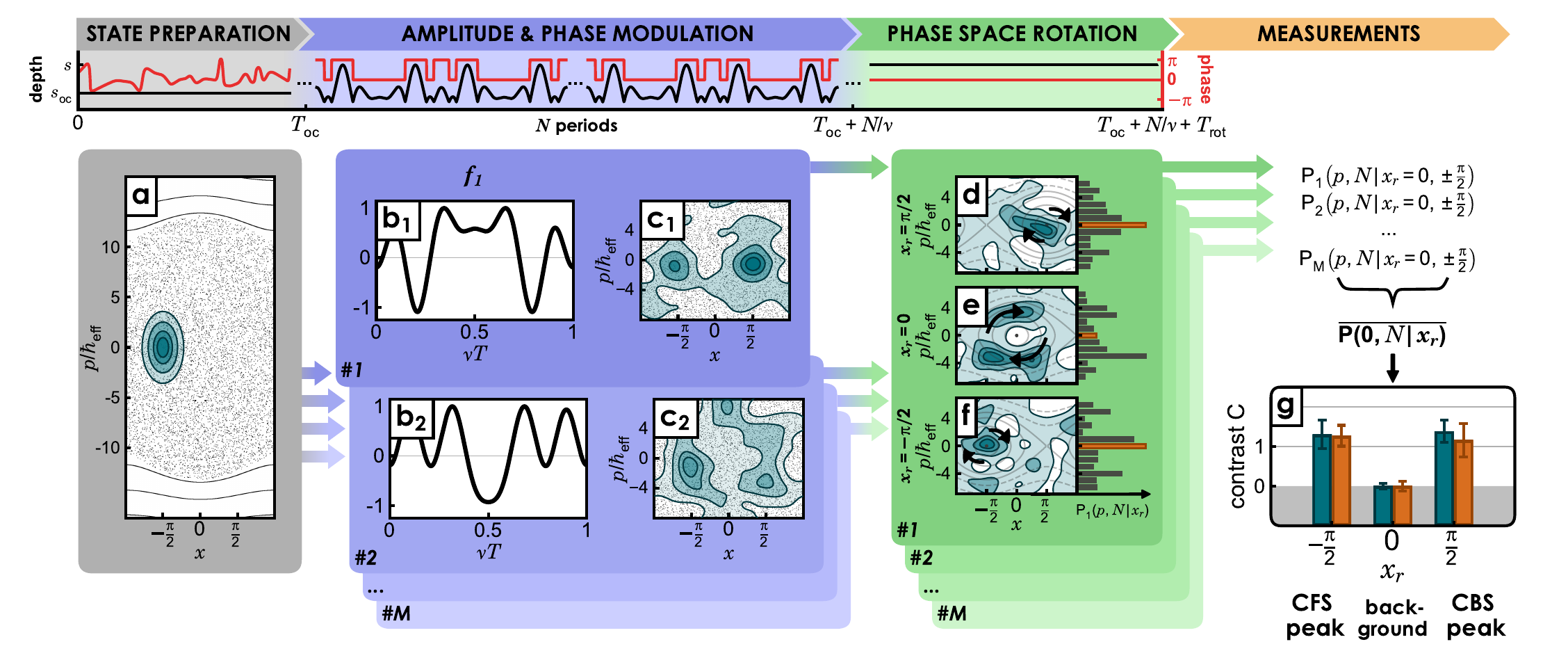}}
    \caption{\textbf{Experimental protocol.} A highly squeezed Gaussian state centered at $x=-x_0=-\pi/2$ is first prepared by a tailored lattice position (phase) modulation, using an optimal control (OC) algorithm for a fixed lattice depth $s_{\rm oc}$ and for a time $T_{\rm oc}$. The Husimi distribution of this initial state is represented in \textbf{a}. The lattice depth is subsequently shaken for $N$ time periods with different periodic modulation functions $f_m$, $1 \leq m\leq M$ (see frames \textbf{b}). The functions have a period $\nu^{-1}$ and contain $N_H$ harmonics with variable phases (see text). 
    After modulation (see \textbf{c}), we perform a phase space rotation, holding the state for a quarter of the period in a static lattice of depth $s$. This transfers the probability density at the center of a lattice site onto the zero momentum component, whose population is measured by imaging the atoms after a time-of-flight (see \textbf{d}, \textbf{e}, \textbf{f}). This procedure is performed three times with three centers of rotation $x=x_r=-\pi/2,0,\pi/2$, set by a shift of the lattice phase, giving the evolutions and momentum populations $P_m(p,N|x_r)$ represented in panels \textbf{d}, \textbf{e}, \textbf{f}, and the bar diagrams on their right. 
    This experiment is repeated $M$ times with different modulation function $f_m$, starting from the same initial state. By averaging all the contributions $P_1$,$...$,$P_M$ we obtain the signal encoding the CFS and CBS peaks and the background in the average amplitude of the zero momentum component. \textbf{(g)} Depending on the modulation regime and symmetries, the CBS or/and CFS peaks emerge from the background (blue, left: numerics, orange, right: experiment). The contrast is defined in \textbf{g} with respect to the background measured for $x_r=0$ (see text).
    Error bars indicate one standard deviation of the mean. Parameters: $s=24.2\pm 0.28$, $M=10$, $N_H=5$, $\nu=35.05$ kHz and $N=18$.}
    \label{fig:Fig2_measurement}
\end{figure*}

\bigskip 

Our experiments realize a \emph{shaken rotor} model:
        \begin{equation}
		\hat{\mathcal{H}} = \frac{\hat{p}^2}{2} - K \cos\left( \hat{x} + \varphi(t) \right)  F(t),
        \label{HSR}
	    \end{equation}
where $\hat{x}$ and $\hat{p}$ are the position and momentum operators respectively, with $-i[\hat{x},\hat{p}]=\hbar_\mathrm{eff}$ an effective Planck constant. 
$K$ is the modulation strength, and $\varphi(t)$ and $F(t)$ describe the periodic position and amplitude modulations of the sinusoidal lattice potential, with period 1. 
The function $F(t)$ sums to 1 over one period of modulation. In the specific case of $\varphi = 0$ and $F = \sum_n \delta(t-n)$, \eqref{HSR} reduces to the celebrated kicked rotor~\cite{ott2002chaos}.

As for the kicked rotor, the shaken rotor Hamiltonian \eqref{HSR} induces fully chaotic classical dynamics within a chaotic sea set by $F(t)$ (see below) for large modulation strengths, $K > 9$ (see Supplementary Information). In the quantum regime, this chaotic dynamics leads to dynamical localization~\cite{ott2002chaos,haake1991quantum,mooreObservationDynamicalLocalization1994e}: the initially diffusive linear increase in momentum variance induced by the shaking is eventually halted through a multiple interference mechanism in momentum space, analogous to Anderson localization~\cite{Casati79}. This non-ergodic behavior gives rise to non-ergodic signatures in the reciprocal space of the localization, \emph{i.e.} in position space~\cite{PhysRevA.95.043626}, which we directly measure for the first time in this work, as illustrated in Figure~\ref{fig:Fig1_schematicillustration}. 

A spatially periodic wavefunction, with an initial narrow distribution within each lattice cell, evolves on average under chaotic dynamics into an almost uniform distribution after a few modulation periods. If the dynamics possess the appropriate symmetry, a CBS peak appears opposite the initial position on the same short timescale. This peak, often discussed for wave scattering in disordered media, has been observed in many contexts \cite{CBSLight1,CBSAcoustic,CBSLight3, CBSLightAtom,CBSPhotonicStructure,CBSLight4, CBSAtom1}. It crucially depends on time-reversal symmetry $\mathsf{T}$ for wave scattering in spatial disorder, and on $\mathsf{PT}$-symmetry, with the transformations $t \rightarrow -t$, $x \rightarrow -x$, and $p \rightarrow p$~\cite{altlandFieldTheoryQuantum1996,PhysRevA.95.043626} in the case of dynamical localization. In the absence of other symmetries, a CFS peak appears at the initial position over a longer timescale, given by the Heisenberg time $t_H$. Depending on the regime, this time scales either as the size of dynamical localization - making it a marker of localization properties~\cite{PhysRevA.95.041602}, or as the size of the classically bounded chaotic region. The CFS peak is therefore more broadly a marker of non-ergodicity in the shaken rotor, compared to the fully chaotic kicked rotor~\cite{ott2002chaos}. The experimentally measured Husimi distributions in Figure~\ref{fig:Fig1_schematicillustration} reveal the finite extent in momentum associated with non-ergodicity, and the characteristic scattering peaks in the position distribution. 
    
The shaken rotor Hamiltonian \eqref{HSR} differs from the kicked rotor in two key aspects. First, the amplitude modulation function is a truncated sum of harmonics:
        \begin{align}
        F(t)=1+2\sum_{n=1}^{N_H} \cos( 2\pi n t+\phi_n),
        \end{align}
where $N_H$ is the number of harmonics, and $\phi_n$ their respective phases. This limits the momentum extension $L_p$ of the chaotic sea, which is proportional to $(2N_H + 1)$. In the quantum regime, the other characteristic length is the localization length $\xi_p$ for dynamical localization, which scales as $K^2$ (see Methods). The model \eqref{HSR} therefore enables to control the localization regime: for parameters where $L_p \gg \xi_p$, non-ergodic properties are induced by localization (\emph{localized} regime), while for $L_p \ll \xi_p$, they correspond to classical confinement within dynamical barriers (\emph{classically bounded} regime). In the latter regime, the dynamics within the boundaries is ergodic, and the growth of the CFS described by random matrix theory (see Supplementary Material). Second, the choice of modulation functions $F$ and $\varphi$ allows us to control the symmetries (such as $\mathsf{T}$ and $\mathsf{PT}$) of the dynamics, leading to enhancement or suppression of the coherent scattering peaks.

Finally, the coherent scattering peaks appear as a statistical effect, requiring averaging over several different classical chaotic dynamics. Here, the use of the shaken rotor with a coherent matter wave provides a straightforward averaging method: for a chosen initial state, localization regime, and symmetries of the dynamics, different modulation functions are available, each leading to different chaotic dynamics (see Methods). Repeating experiments with several modulation functions, we obtain the averaged scattering peaks in the chosen regime. This allows us in particular to probe the coherent scattering peaks in the presence of $\mathsf{T}$ symmetry, a regime unachievable with the average over initial conditions usually performed in previous kicked rotor experiments~\cite{mooreObservationDynamicalLocalization1994e}.

\begin{figure*}[t]
    \centering
    \includegraphics[width=\textwidth]{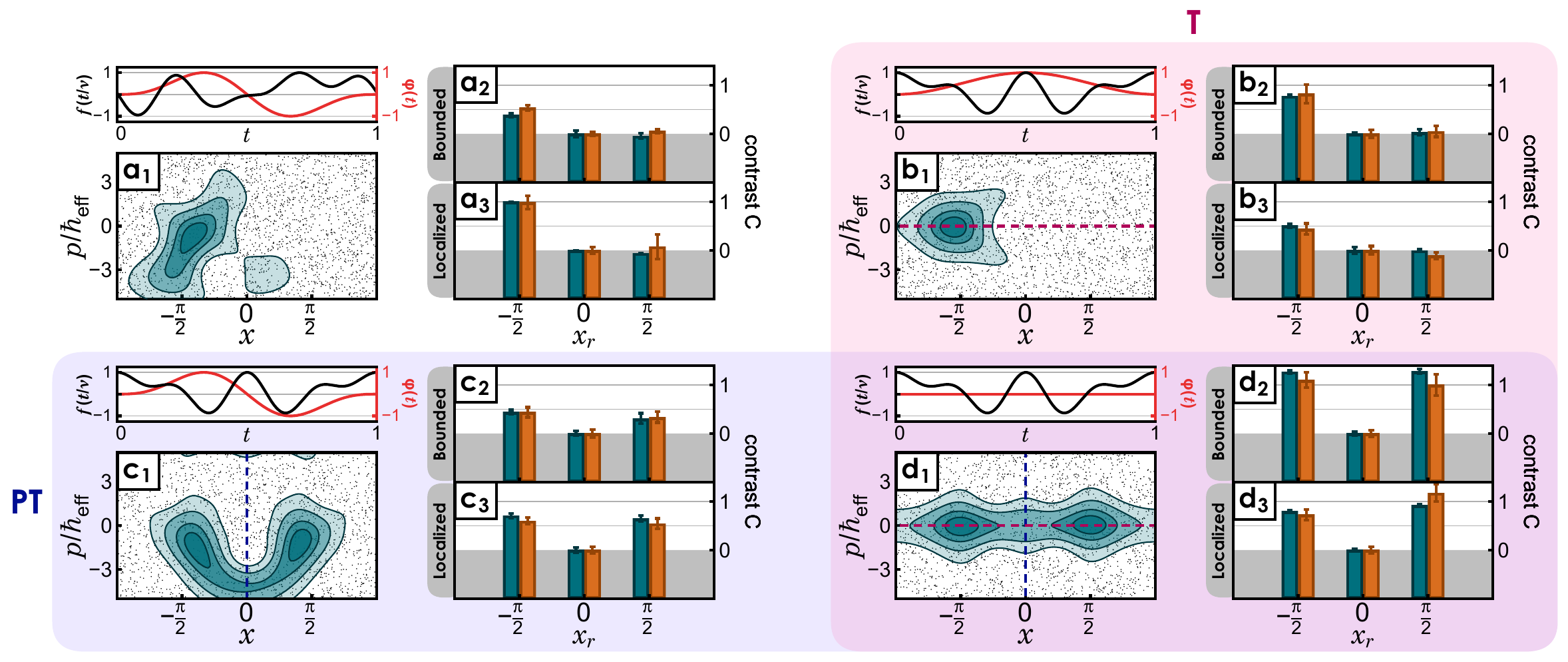}
    \caption{Top row in each panel (\textbf{a-d}): Shape of periodic modulations of lattice depth (black) and position (red) corresponding to the symmetry regime (none, $\mathsf{T}$, $\mathsf{PT}$, $\mathsf{P+T}$). (\textbf{a\textsubscript{1}–d\textsubscript{1}}) Husimi representations of the maximum overlap Floquet eigenstate (see text) in each regime of symmetry. (\textbf{a\textsubscript{2,3}–d\textsubscript{2,3}}) Experimental (orange) and numerical (blue) contrasts obtained for different symmetries of the modulation, and in two different localization regimes:  bounded $\xi_{\rm loc}/L\gtrsim 1$, and localized $\xi_{\rm loc}/L\lesssim 0.1$, where $\xi_{\rm loc}=\xi_p/\hbar_\mathrm{eff}$ is the localization length and $L=L_p/\hbar_\mathrm{eff}$ the extension of the chaotic sea, in units of $\hbar_\mathrm{eff}$. The signals are obtained using the method presented in Fig.~\ref{fig:Fig2_measurement} with an additional average over 3 modulations periods ($N=[15,16,17]$ and $M=10$ if not otherwise stated).  Parameters: ($\mathbf{a_2}$) $L=9.24\pm0.16$, $\xi_{\rm loc}=42\pm26$, $M=17$, ($\mathbf{a_3}$) $L=48.56\pm0.24$, $\xi_{\rm loc}=4.47\pm1.43$, $N=[6,7,8]$, ($\mathbf{b_2}$) $L=9.04\pm0.10$, $\xi_{\rm loc}=29.5\pm 16$, ($\mathbf{b_3}$) $L=23.96\pm0.38$, $\xi_{\rm loc}=1.52\pm 0.35$, ($\mathbf{c_2}$) $L=9.90\pm0.10$, $\xi_{\rm loc}=31\pm16$, ($\mathbf{c_3}$) $L=62.70\pm0.58$, $\xi_{\rm loc}=1.9\pm0.73$,  ($\mathbf{d_2}$) $L=8.42\pm0.32$, $\xi_{\rm loc}=20.1\pm 8.5$, ($\mathbf{d_3}$) $L=25.78\pm0.42$, $\xi_{\rm loc}=0.77\pm 0.05$.
    }
    \label{fig:Fig3_table&Husimi}
\end{figure*}

\bigskip

Our experiments start with a BEC of about $5\cdot10^5$
rubidium-87 atoms produced in a hybrid trap (see Methods) and placed in a 1D optical lattice potential:
    \begin{equation}
    V(X,T)=-\frac{s}{2}f(T)E_L \cos(2\pi \frac{X}{d}+\varphi(T))
    \label{eq:latticePotential}
    \end{equation}
 with $d=532\,{\rm nm}$ the lattice spacing, $E_L=h^2/(2md^2)=h\nu_L$ its energy scale, $h$ the Planck constant and $m$ the atomic mass. The optical lattice is produced by interference of two far-detuned counter-propagating beams derived from the same laser, whose amplitude and phase are controlled in time by acousto-optic modulators (AOM). We vary the depth $sf(T)$ of the lattice potential, where $|f(T)|\leq 1$ and $s$ is the maximum achievable depth, and its position $\varphi(T)$.

For a periodic modulation of the lattice with frequency $\nu$, we recover the Hamiltonian \eqref{HSR} with dimensionless variables $t=\nu T$, $x=2\pi X/d$, and $p=2\pi P/(md\nu)$, modulation function $f(T)=F(\nu T)/\max(|F|)$ and modulation strength $K=4\pi^2s(\nu_L/\nu)^2/\max(|F|)$. The tunable effective Planck constant is $\hbar_\mathrm{eff}=4\pi(\nu_L/\nu)$. %

A single experimental sequence comprises three steps, sketched in Figure~\ref{fig:Fig2_measurement}. The first step prepares the initial quantum state in the lattice:
the BEC is adiabatically loaded into the ground state of a static lattice with a depth of $s \times f(T = 0) \simeq 5$. In the lattice band structure, this ground state belongs to the subspace of zero quasi-momentum, preserved under modulation. Therefore, the subsequent evolution of the state can be expressed as a superposition of plane waves with discrete momenta $p=\ell \hbar_\mathrm{eff}$. Throughout the dynamics, the state can be characterized by a measurement of the average populations in these momentum components, performed by absorption imaging of the BEC released from the trap after a time-of-flight.

The ground state is then transformed into a periodic state with a squeezed Gaussian position distribution in each lattice site, centered on $x=-x_0=-\pi/2$, and of width $\Delta x/(2\pi)\simeq 4\%$ (correspondingly $\Delta p/\hbar_{\rm eff}\simeq 1.9$) for experiments presented here (see Fig.~\ref{fig:Fig2_measurement} {\bf a}). Such a narrow Gaussian distribution cannot be achieved through adiabatic loading, as it would require extraordinary lattice depths ($s \simeq 200$). We therefore perform quantum state preparation with a modulation of the lattice phase $\varphi(T)$ derived from a quantum optimal control algorithm~\cite{dupont_quantum_2021}. This state preparation modulation is performed in the initial fixed-depth lattice with $s_{oc}=s\times f\simeq5$ and allows us to produce the peaked initial state in a typical duration $T_{oc} \simeq 100\,\mu$s, with high fidelity  (see Fig.~\ref{fig:Fig1_schematicillustration} {\bf a}, Fig.~\ref{fig:Fig2_measurement} {\bf a} and Methods).

The precise choice and preparation of the initial state are crucial to our measurements. The width of the initial state results from a compromise: a too broad position distribution leads to reduced CFS and CBS contrasts~\cite{PhysRevA.95.043626}, but the initial state also has to primarily evolve inside the chaotic region, implying a momentum extension smaller than $L_p$.

The second and main step performs the chaotic dynamics, through a choice of periodic modulation functions $\{F(t),\varphi(t)\}$ with given symmetries. 
Negative values of $F(t)$ are implemented via a sudden $\pi$-shift of the lattice phase $\varphi$ when the lattice amplitude reaches zero (see Figure~\ref{fig:Fig2_measurement}).
After a given number $N$ of modulation periods, the system has undergone chaotic dynamics, and displays on average a spreading over the classical chaotic sea, with the coherent scattering peaks superimposed (see Fig.~\ref{fig:Fig1_schematicillustration}). 

The final step involves dynamics in a regular phase space: the state is held in a static lattice with maximal depth $s$ for a duration close to a quarter-period at the lattice well frequency $T_\mathrm{rot}\simeq(\sqrt{s} \nu_L)^{-1}/4$. This amounts to a $\pi/2$-rotation in phase space around the bottom of the well, transferring the position distribution into momentum space~\cite{PhysRevA.95.043626}. Specifically, by positioning the expected coherent peaks at the center of the wells through a sudden position shift $x_r = \pm \pi/2$ before the rotation, we convert them into a peak in the momentum distribution $|c_\ell|^2$ at $\ell = p / \hbar_\mathrm{eff}  =0$ (see Figure~\ref{fig:Fig2_measurement} {\bf d-f}). The measured probability $P(p = 0, N | x_r = \pm \pi/2)$ can then be compared to the value for $x_r = 0$, which reflects the background. 

This final phase space rotation must accurately convert the position distribution peak into a peak at momentum $p = 0$. This turns out to be the primary limit on the momentum extension of the initial squeezed state. Initial state preparation and phase space rotation parameters were optimized in light of these constraints (see Methods).

Each choice of modulation functions $\{F(t),\varphi(t)\}$ thus implies at least three full experimental sequences. Moreover, we perform an average over disorder, \emph{i.e.} over different modulation functions with given parameters $(K,\hbar_\mathrm{eff})$ and chosen symmetries: 
for each configuration considered, we average over $M=10$ amplitude modulation functions $F(t)$, with a fixed phase modulation $\varphi(t)$ setting the symmetries of the dynamics. The averaged probability $\overline{P(p=0,N|x_r=\pm\pi/2)}$ measured near zero momentum after phase space rotation and time-of-flight, compared to the value obtained for $x_r=0$, constitutes a direct measurement of the contrast of the CFS and CBS peaks, defined as $\mathsf{C}_{CBS,CFS}=\overline{P(0,N|x_r=\pm\pi/2)}/\overline{P(0,N|x_r=0)}-1$, as shown in Figure~\ref{fig:Fig2_measurement} {\bf g}. Our experimental results are compared to extensive numerical simulations of the dynamics through the time-dependent Schrödinger equation (TDSE - see Methods).

\begin{figure*}[ht!]
    \centering
    \includegraphics[width=\textwidth]{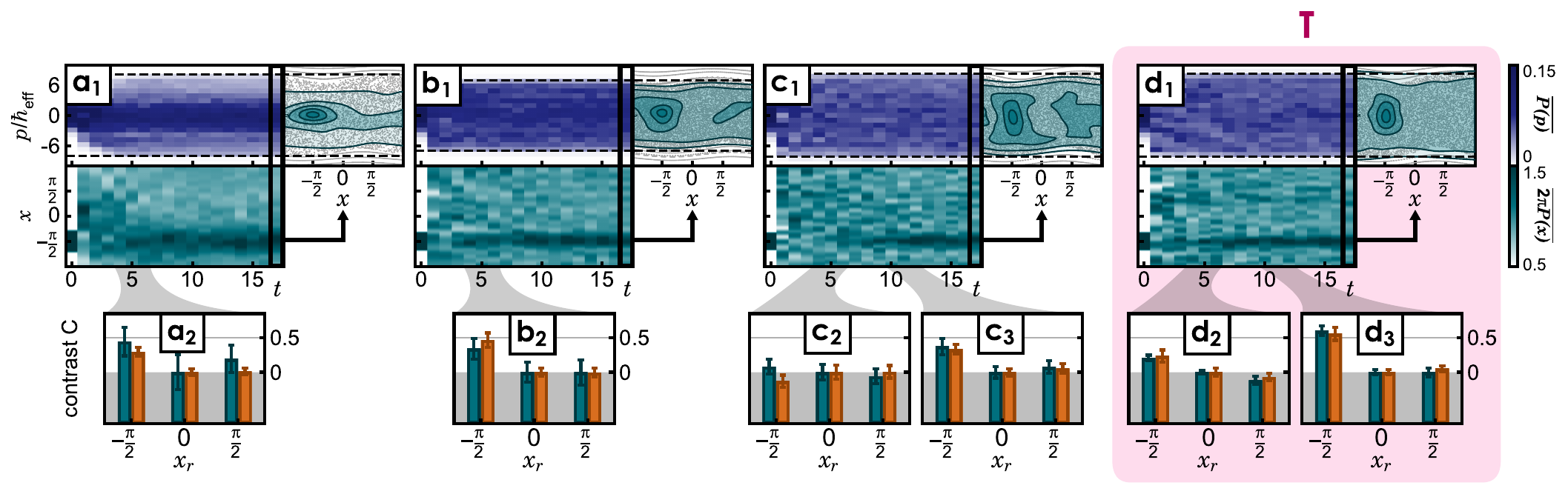}
    \caption{ \textbf{
    CFS apparition timescale for various localization regimes and chaotic sea sizes $L$, and impact of $\mathsf{T}$-symmetry.} 
Top panels show the averaged momentum distribution $\overline{P(p)}$ as a function of time, with the Husimi representation of the final state to the right. Bottom panels show the corresponding evolution of the averaged spatial density $\overline{P(x)}$. These numerical results are computed in a localized regime with large chaotic sea bounded by the dashed lines ($\mathbf{a_1}$), a classically bounded regime with a ``small'' chaotic sea ($\mathbf{b_1}$) and a bounded regime with a ``large'' chaotic sea ($\mathbf{c_1}-\mathbf{d_1}$). In $\mathbf{d_1}$, the modulation is chosen so that dynamics are $\mathsf{T}$-symmetry invariant, while all symmetries are broken for $\mathbf{a_1-c_1}$.
Experimental and numerical signals (obtained as per Fig.~\ref{fig:Fig2_measurement}) are shown at a short modulation time ($N=[4,5,6]$) for all cases ($\mathbf{a_2}-\mathbf{d_2}$), and at a longer time for cases $\textbf{c-d}$ ($N=[10,11,12]$, $\mathbf{c_3-d_3}$), showing the longer timescale of evolution of the CFS peak for larger classical bounds. The addition of $\mathsf{T}$-symmetry in $\textbf{d}$ strikingly impacts the CFS in the bounded regime shared by $\mathbf{c-d}$: the contrast is nonzero at short times, and saturates to an enhanced value (see text). 
Parameters: (\textbf{a}) $L=15.52\pm 0.34$, $\xi_{\rm loc}=5.97\pm2.8$, (\textbf{b}) $L=12.8\pm0.46$, $\xi_{\rm loc}=18.5\pm7.2$, (\textbf{c}) $L=15.9\pm0.24$, $\xi_{\rm loc}=52.5\pm26$, (\textbf{d}) $L=15.6\pm0.19$, $\xi_{\rm loc}=51\pm 31$.}
    \label{fig:Fig4_CFSdynamics}
\end{figure*}

We have applied our measurement protocol in the localized and classically bounded regimes of the shaken rotor, varying the symmetry regime. 
The tailoring of symmetries through modulation functions is illustrated in Figure~\ref{fig:Fig3_table&Husimi}. 
A key symmetry of interest is $\mathsf{PT}$ symmetry (Fig.~\ref{fig:Fig3_table&Husimi} {\bf c}), achieved through a combination of time-symmetric amplitude modulation, $F(1-t) = F(t)$, and anti-symmetric position modulation, $\varphi(1-t) = -\varphi(t)$. Symmetric amplitude and position modulation realize the usual time-reversal $\mathsf{T}$ symmetry (Fig.~\ref{fig:Fig3_table&Husimi} {\bf b}), and we can combine both symmetries, with symmetric $F$ and $\varphi = 0$, to obtain dynamics that are both $\mathsf{P}$ and $\mathsf{T}$ invariant (Fig.~\ref{fig:Fig3_table&Husimi} {\bf d}). The periodic phase modulation is chosen with a minimum number of harmonics while ensuring $\mathrm{d}\varphi/\mathrm{d}t(t=0)=0$, to guarantee a continuous variation of momentum in the lattice. The amplitude modulation $F(t)$ is symmetric for a random choice of $\phi_n = 0$ or $\pi$ for each harmonic phase, while $\phi_n$ can be chosen randomly between $0$ and $2\pi$ to break the symmetry.
The symmetries of the dynamics are reflected in the structure of the eigenstates of the one-period evolution operator, or Floquet states, as highlighted in Figure~\ref{fig:Fig3_table&Husimi} {$\mathrm{\mathbf{a_1-d_1}}$}, which shows the Husimi distribution of the Floquet state with maximum overlap to the initial peaked state in $x=-x_0$, for each symmetry regime. 
    
The CBS and CFS contrasts in Figure~\ref{fig:Fig3_table&Husimi} constitute the main result of this work, and vividly illustrate the impact of both symmetry and localization regimes on these coherent signatures of non-ergodicity. In all regimes (Fig.~\ref{fig:Fig3_table&Husimi} {\bf a-d}) a CFS peak is present, demonstrating that it is a robust marker of non-ergodicity, be it from dynamical localization over a length $\xi_p$, or box-constrained dynamics with extension $L_p$. The CBS peak, which relies on the interference of symmetric trajectories, is only observed when the appropriate $\mathsf{PT}$ symmetry is present (Fig.~\ref{fig:Fig3_table&Husimi} {\bf b} and {\bf d}), where it mirrors the CFS peak. 
We also observe the enhancement of the CFS when $\mathsf{T}$-symmetry is present, in the bounded regime (Fig.~\ref{fig:Fig3_table&Husimi} $\mathrm{\mathbf{b_2}}$), with a clear increase of the CFS contrast. 
This CFS enhancement arises from constructive interferences that occur in the presence of $\mathsf{T}$-symmetry: chaotic trajectories originating from and returning to the original position interfere with time-reversed counterparts, increasing the probability at the peak position~\cite{Hebraud25}.
In the localized regime where $L_p \gg \xi_p$, this enhancement is blurred by the finite width $\Delta x \gtrsim \hbar_{\rm eff} / L_p $ of the initial state~\cite{PhysRevA.95.043626}. The addition of $\mathsf{T}$-symmetry thus does not lead to a measurable increase of the CFS contrast in this regime (Fig.~\ref{fig:Fig3_table&Husimi} $\mathrm{\mathbf{c_2}}$).
Combining both symmetries, we recover the CBS peak alongside the CFS peak ($\mathsf{PT}-$symmetry), with an additional doubling of both CBS and CFS contrasts in the bounded regime due to $\mathsf{T}-$symmetry (Fig.~\ref{fig:Fig3_table&Husimi} $\mathrm{\mathbf{d_2}}$). 

All measurements in Figure~\ref{fig:Fig3_table&Husimi} are compared to TDSE simulations, showing good agreement with the observed CBS and CFS contrasts. While the time-dependent growth of the CFS peak is known to encode the spectral form factor and reveal key symmetries, this is also true of the asymptotic contrast, as shown by the impact of $\mathsf{T}$ symmetry. 

The tomographic approach from which we extract the signal of the coherent scattering peaks of Figure~\ref{fig:Fig3_table&Husimi} can be extended to perform full state characterization~\cite{dupont_phase-space_2023}. By modifying the final step in Figure~\ref{fig:Fig2_measurement}, we can instead record the momentum distribution for several holding times in the static lattice of depth $s$. This dynamical evolution provides a complete tomography of the final state of the system, from which we can obtain a maximum likelihood state estimate, $\hat{\rho}_{ML}$ (see Methods). With this state estimate, the phase space Husimi quasi-probability distribution can be readily computed. This was done for the initial state shown in Figure~\ref{fig:Fig1_schematicillustration}~{\bf a}.
For the final states, averaging state reconstruction results over the $M$ modulation functions provides a disorder-averaged state estimate, and the phase space distributions shown in Fig.~\ref{fig:Fig1_schematicillustration}~{\bf b,c} reveal the coherent scattering peaks. This gives access to the width of the CFS peak, which is expected to scale inversely with the localization or confinement length. From our full state tomography, we extract a full width at half maximum (FWHM) of the CFS peak of $0.64\pm 0.12$ (resp. $0.52\pm 0.07$) in dimensionless position units, for the data of Fig.~\ref{fig:Fig1_schematicillustration}~{\bf b} (resp. {\bf c}), in good agreement with the numerical prediction $0.66\pm0.06$ (resp. $0.54\pm 0.04$). This also agrees with an estimate of the width from the combination of the expected CFS width from the classical bounds and the width of the prepared initial state, $\rm FWHM\simeq0.64$ for $L=9$.  Such an observable does not exist when probing CFS-related contributions in a synthetic system such as~\cite{hainaut2018controlling}.

Finally, we experimentally demonstrate that the CFS is a \emph{quantitative} marker of non-ergodicity, by investigating its growth dynamics, shown in Figure~\ref{fig:Fig4_CFSdynamics}. In the absence of other symmetries, the timescale $t_H$ for appearance of the CFS is that of localization, either from dynamical localization ($t_H\sim\xi_{loc}\equiv\xi_p/\hbar_{\rm eff}$) or from the bounded chaotic dynamics ($t_H\sim L\equiv L/\hbar_{\rm eff}$). We highlight this link by tuning both $\xi_{loc}$ and $L$, with results shown in Figure~\ref{fig:Fig4_CFSdynamics}. In Fig.~\ref{fig:Fig4_CFSdynamics} {\bf a-c} no $\mathsf{PT}$ symmetry is present, and only the CFS peak is expected. In Fig.~\ref{fig:Fig4_CFSdynamics} $\mathrm{\mathbf{a_2}}$ and $\mathrm{\mathbf{b_2}}$, the CFS is clearly visible in measurements after approximately 5 periods of modulation. This non-ergodic signature is, however, of two distinct origins, (exponential) dynamical localization with $\xi^{(a)}_{loc}\ll L^{(a)}$, or bounded diffusion within a limited-size box $L^{(b)}$, as illustrated by numerically computed evolutions (Fig.~\ref{fig:Fig4_CFSdynamics} $\mathrm{\mathbf{a_1}}$ and $\mathrm{\mathbf{b_1}}$). In Fig.~\ref{fig:Fig4_CFSdynamics} {\bf c}, modulation parameters are chosen such that the dynamics is also essentially a bounded diffusion within a chaotic sea of size $L^{(c)}\simeq L^{(a)}>L^{(b)}$. The growth of the CFS is delayed, as expected, compared to the other measurements, and it becomes visible after about 11 periods of modulation. Finally in Fig.~\ref{fig:Fig4_CFSdynamics} {\bf d}, we highlight the changes in CFS dynamics brought by the $\mathsf{T}$ symmetry, in the same non-ergodicity regime as Fig.~\ref{fig:Fig4_CFSdynamics} {\bf c}, of classically bounded diffusion with $L^{(d)}\simeq L^{(c)}$: a non-zero CFS contrast is quickly established at short times, with the contrast further rising to an enhanced value compared to Fig.~\ref{fig:Fig4_CFSdynamics}~{\bf c} over the localization time. Simultaneously, no CBS peak develops, as $\mathsf{PT}$-symmetry remains broken. The measurements of this unusual behavior, predicted in the kicked rotor in this regime of symmetry~\cite{Hebraud25}, agree well with numerical simulations. These results manifest the critical interplay between the characteristic sizes $\xi_\mathrm{loc}$ and $L$ and the symmetries in setting the CFS dynamics, which conversely confirms the CFS  as a quantitative marker of the non-ergodicity and symmetry properties of the system.

In conclusion, we have demonstrated a new method for the measurement of the CFS using cold atoms in a shaken rotor potential. Critically, our system allows for a genuine average over chaotic dynamics, while retaining control over the effective system size, localization length and symmetries of the dynamics. With this method, we have realized the first direct measurement of the CFS peak appearing in position space. Varying the symmetry regimes, we have highlighted their role in the appearance and contrast of both scattering peaks, and evidenced a enhancement of contrast in the presence of time-reversal symmetry. We have measured the dependence of the CFS growth on the controllable characteristic lengths and symmetry, establishing it as a quantitative marker of localization properties.
Given the ubiquitous nature of the CFS, our results open new avenues for probing the coherent signatures of non-ergodicity in quantum many-body chaos and localization.

\paragraph*{Acknowledgements}
We are grateful to the late D. Delande for many insightful and enriching discussions. We thank Calcul en Midi-Pyr\'en\'ees (CALMIP) for computational resources. This work was supported by the ANR projects QuCoBEC (ANR-22-CE47-0008)  Gladys (ANR-19-CE30-0013), QUTISYM (ANR-23-PETQ-0002) and ManyBodyNet, the EUR Grant NanoX No. ANR-17-EURE-0009, by the Singapore Ministry of Education Academic Research Funds Tier II (WBS No. A-8001527-
02-00 and A-8002396-00-00) and the ERC Grant LATIS.

	\bibliography{references.bib}

@article{CBSAcoustic,
  title = {Observation of acoustic coherent backscattering},
  author = {Sakai, Keiji and Yamamoto, Ken and Takagi, Kenshiro},
  journal = {Phys. Rev. B},
  volume = {56},
  issue = {17},
  pages = {10930--10933},
  numpages = {0},
  year = {1997},
  month = {Nov},
  publisher = {American Physical Society},
  doi = {10.1103/PhysRevB.56.10930},
  url = {https://link.aps.org/doi/10.1103/PhysRevB.56.10930}
}

@article{CBSLight1,
  title = {Weak Localization and Coherent Backscattering of Photons in Disordered Media},
  author = {Wolf, Pierre-Etienne and Maret, Georg},
  journal = {Phys. Rev. Lett.},
  volume = {55},
  issue = {24},
  pages = {2696--2699},
  numpages = {0},
  year = {1985},
  month = {Dec},
  publisher = {American Physical Society},
  doi = {10.1103/PhysRevLett.55.2696},
  url = {https://link.aps.org/doi/10.1103/PhysRevLett.55.2696}
}

@article{CBSLight3,
	author = {Wiersma, Diederik S. and Bartolini, Paolo and Lagendijk, Ad and Righini, Roberto},
	journal = {Nature},
	number = {6661},
	pages = {671--673},
	title = {Localization of light in a disordered medium},
	volume = {390},
    doi={10.1038/37757},
	year = {1997}}

@article{CBSLight4,
  title = {Control of coherent backscattering by breaking optical reciprocity},
  author = {Bromberg, Y. and Redding, B. and Popoff, S. M. and Cao, H.},
  journal = {Phys. Rev. A},
  volume = {93},
  issue = {2},
  pages = {023826},
  numpages = {6},
  year = {2016},
  month = {Feb},
  publisher = {American Physical Society},
  doi = {10.1103/PhysRevA.93.023826},
  url = {https://link.aps.org/doi/10.1103/PhysRevA.93.023826}
}

@article{CBSPhotonicStructure,
  title = {Anomalous Coherent Backscattering of Light from Opal Photonic Crystals},
  author = {Huang, J. and Eradat, N. and Raikh, M. E. and Vardeny, Z. V. and Zakhidov, A. A. and Baughman, R. H.},
  journal = {Phys. Rev. Lett.},
  volume = {86},
  issue = {21},
  pages = {4815--4818},
  numpages = {0},
  year = {2001},
  month = {May},
  publisher = {American Physical Society},
  doi = {10.1103/PhysRevLett.86.4815},
  url = {https://link.aps.org/doi/10.1103/PhysRevLett.86.4815}
}

@article{CBSAtom1,
  title = {Coherent Backscattering of Ultracold Atoms},
  author = {Jendrzejewski, F. and M\"uller, K. and Richard, J. and Date, A. and Plisson, T. and Bouyer, P. and Aspect, A. and Josse, V.},
  journal = {Phys. Rev. Lett.},
  volume = {109},
  issue = {19},
  pages = {195302},
  numpages = {5},
  year = {2012},
  month = {Nov},
  publisher = {American Physical Society},
  doi = {10.1103/PhysRevLett.109.195302},
  url = {https://link.aps.org/doi/10.1103/PhysRevLett.109.195302}
}

@article{CBSLightAtom,
  title = {Coherent Backscattering of Light by Cold Atoms},
  author = {Labeyrie, G. and de Tomasi, F. and Bernard, J.-C. and M\"uller, C. A. and Miniatura, C. and Kaiser, R.},
  journal = {Phys. Rev. Lett.},
  volume = {83},
  issue = {25},
  pages = {5266--5269},
  numpages = {0},
  year = {1999},
  month = {Dec},
  publisher = {American Physical Society},
  doi = {10.1103/PhysRevLett.83.5266},
  url = {https://link.aps.org/doi/10.1103/PhysRevLett.83.5266}
}

@book{ott2002chaos,
  title={{Chaos in Dynamical Systems}},
  author={Ott, Edward},
  year={2002},
  publisher={{Cambridge University Press}},
  doi={10.1017/CBO9780511803260}
}

@inproceedings{Casati79,
    author = {Casati,G. and Chirikov, B. V. and Izraelev F. M.  and Ford, J.},
    title = {Stochastic behavior of a quantum pendulum under a periodic perturbation},
    booktitle = {Stochastic Behavior in Classical and Quantum Hamiltonian Systems},
    year = {1979},
    series = {Lecture Notes in Physics},
    volume = {93},
    editor = {Casati, G. and Ford, J.} ,
    publisher={Springer, Berlin},
    doi={10.1007/BFb0021732}
}

@article{d2016quantum,
  title={From quantum chaos and eigenstate thermalization to statistical mechanics and thermodynamics},
  author={D'Alessio, Luca and Kafri, Yariv and Polkovnikov, Anatoli and Rigol, Marcos},
  journal={Advances in Physics},
  volume={65},
  number={3},
  pages={239--362},
  year={2016},
  publisher={Taylor \& Francis},
  doi={10.1080/00018732.2016.1198134}
}

@misc{Madani24,
    title={},
    author={Madani, F. and Denis, M. and Szriftgiser, P. and Garreau, J.-C. and Rançon, A. and Chicireanu, R.},
    archivePrefix = {arXiv},
    eprint = {2402.06573},
    year = {2024}
}

@article{Gogolin_2016,
doi = {10.1088/0034-4885/79/5/056001},
url = {https://dx.doi.org/10.1088/0034-4885/79/5/056001},
year = {2016},
month = {apr},
publisher = {IOP Publishing},
volume = {79},
number = {5},
pages = {056001},
author = {Christian Gogolin and Jens Eisert},
title = {Equilibration, thermalisation, and the emergence of statistical mechanics in closed quantum systems},
journal = {Rep. Prog. Phys.},
}

@article{ueda2020quantum,
  title={Quantum equilibration, thermalization and prethermalization in ultracold atoms},
  author={Ueda, Masahito},
  journal={Nature Reviews Physics},
  volume={2},
  number={12},
  pages={669--681},
  year={2020},
  publisher={Nature Publishing Group UK London},
  doi={10.1038/s42254-020-0237-x}
}

@article{PhysRev.109.1492,
  title = {Absence of Diffusion in Certain Random Lattices},
  author = {Anderson, P. W.},
  journal = {Phys. Rev.},
  volume = {109},
  issue = {5},
  pages = {1492--1505},
  numpages = {0},
  year = {1958},
  month = {Mar},
  publisher = {American Physical Society},
  doi = {10.1103/PhysRev.109.1492},
  url = {https://link.aps.org/doi/10.1103/PhysRev.109.1492}
}

@article{RevModPhys.80.1355,
  title = {Anderson transitions},
  author = {Evers, Ferdinand and Mirlin, Alexander D.},
  journal = {Rev. Mod. Phys.},
  volume = {80},
  issue = {4},
  pages = {1355--1417},
  numpages = {0},
  year = {2008},
  month = {Oct},
  publisher = {American Physical Society},
  doi = {10.1103/RevModPhys.80.1355},
  url = {https://link.aps.org/doi/10.1103/RevModPhys.80.1355}
}

@book{anderson201050,
  title={50 Years of Anderson Localization},
  doi={10.1142/7663},
  editor={Abrahams, E.},
  publisher={World Scientific},
  year={2010}
}

@article{RevModPhys.91.021001,
  title = {Colloquium: Many-body localization, thermalization, and entanglement},
  author = {Abanin, Dmitry A. and Altman, Ehud and Bloch, Immanuel and Serbyn, Maksym},
  journal = {Rev. Mod. Phys.},
  volume = {91},
  issue = {2},
  pages = {021001},
  numpages = {26},
  year = {2019},
  month = {May},
  publisher = {American Physical Society},
  doi = {10.1103/RevModPhys.91.021001},
  url = {https://link.aps.org/doi/10.1103/RevModPhys.91.021001}
}

@article{sierant2025many,
  author={Sierant, Piotr and Lewenstein, Maciej and Scardicchio, Antonello and Vidmar, Lev and Zakrzewski, Jakub},
  title={Many-body localization in the age of classical computing},
  journal={Rep. Prog. Phys.},
  volume={88},
  pages={026502},
  doi={10.1088/1361-6633/ad9756},
  url={https://iopscience.iop.org/article/10.1088/1361-6633/ad9756},
  year={2025}
}

@article{alet2018many,
  title={Many-body localization: An introduction and selected topics},
  author={Alet, Fabien and Laflorencie, Nicolas},
  journal={Comptes Rendus Physique},
  volume={19},
  number={6},
  pages={498--525},
  year={2018},
   doi={10.1016/j.crhy.2018.03.003},
  publisher={Elsevier}
}

@article{serbyn2021quantum,
  title={Quantum many-body scars and weak breaking of ergodicity},
  author={Serbyn, Maksym and Abanin, Dmitry A and Papi{\'c}, Zlatko},
  journal={Nature Physics},
  volume={17},
  number={6},
  pages={675--685},
  year={2021},
  doi={10.1038/s41567-021-01230-2},
  publisher={Nature Publishing Group UK London}
}

@article{bohigas1993manifestations,
  title={Manifestations of classical phase space structures in quantum mechanics},
  author={Bohigas, Oriol and Tomsovic, Steven and Ullmo, Denis},
  journal={Physics Reports},
  volume={223},
  number={2},
  pages={43--133},
  year={1993},
  publisher={Elsevier},
  doi={10.1016/0370-1573(93)90109-Q}
}

@article{santhanam2022quantum,
  title={Quantum kicked rotor and its variants: Chaos, localization and beyond},
  author={Santhanam, MS and Paul, Sanku and Kannan, J Bharathi},
  journal={Physics Reports},
  volume={956},
  pages={1--87},
  year={2022},
  doi={10.1016/j.physrep.2022.01.002},
  publisher={Elsevier}
}

@article{tikhonov2021anderson,
  title={From Anderson localization on random regular graphs to many-body localization},
  author={Tikhonov, Konstantin S and Mirlin, Alexander D},
  journal={Annals of Physics},
  volume={435},
  pages={168525},
  year={2021},
  doi={10.1016/j.aop.2021.168525},
  publisher={Elsevier}
}

@book{akkermans2007mesoscopic,
  title={Mesoscopic physics of electrons and photons},
  author={Akkermans, Eric and Montambaux, Gilles},
  year={2007},
  doi={10.1017/CBO9780511618833},
  publisher={Cambridge University Press}
}

@article{hainaut2017ERO,
  title = {Return to the Origin as a Probe of Atomic Phase Coherence},
  author = {Hainaut, C. and Manai, I. and Chicireanu, R. and Cl\'ement, J.-F. and Zemmouri, S. and Garreau, J.-C and Szriftgiser, P. and Lemari\'e, G. and Cherroret, N. and Delande, D.},
  journal = {Phys. Rev. Lett.},
  volume = {118},
  issue = {18},
  pages = {184101},
  year = {2017},
  doi = {10.1103/PhysRevLett.118.184101},
  url = {https://link.aps.org/doi/10.1103/PhysRevLett.118.184101}
}

@article{hainaut2018controlling,
  title={Controlling symmetry and localization with an artificial gauge field in a disordered quantum system},
  author={Hainaut, Cl{\'e}ment and Manai, Isam and Cl{\'e}ment, Jean-Fran{\c{c}}ois and Garreau, Jean Claude and Szriftgiser, Pascal and Lemari{\'e}, Gabriel and Cherroret, Nicolas and Delande, Dominique and Chicireanu, Radu},
  journal={Nat. Commun.},
  volume={9},
  number={1},
  pages={1382},
  year={2018},
  doi={10.1038/s41467-018-03481-9},
  publisher={Nature Publishing Group UK London}
}

@article{Chabe08,
  title = {Experimental Observation of the Anderson Metal-Insulator Transition with Atomic Matter Waves},
  author = {Chab\'e, Julien and Lemari\'e, Gabriel and Gr\'emaud, Beno\^{\i}t and Delande, Dominique and Szriftgiser, Pascal and Garreau, Jean Claude},
  journal = {Phys. Rev. Lett.},
  volume = {101},
  issue = {25},
  pages = {255702},
  year = {2008},
  doi = {10.1103/PhysRevLett.101.255702},
  url = {https://link.aps.org/doi/10.1103/PhysRevLett.101.255702}
}

@article{Billy08,
  title = {Direct observation of Anderson localization of matter waves in a controlled disorder},
  author = {Billy, J. and Josse, V. and Zuo, Z. and Bernard, A. and Hambrecht, B. and Lugan, P. and Clément, D. and Sanchez-Palencia, L. and Bouyer, P. and Aspect, A.},
  journal = {Nature},
  volume = {453},
  pages = {891–894},
  year = {2008},
  doi = {10.1038/nature07000},
  url = {https://www.nature.com/articles/nature07000}
}

@article{Sajjad22,
  title = {Observation of the Quantum Boomerang Effect},
  author = {Sajjad, Roshan and Tanlimco, Jeremy L. and Mas, Hector and Cao, Alec and Nolasco-Martinez, Eber and Simmons, Ethan Q. and Santos, Fl\'avio L. N. and Vignolo, Patrizia and Macr\`{\i}, Tommaso and Weld, David M.},
  journal = {Phys. Rev. X},
  volume = {12},
  pages = {011035},
  year = {2022},
  doi = {10.1103/PhysRevX.12.011035},
  url = {https://link.aps.org/doi/10.1103/PhysRevX.12.011035}
}

@book{haake1991quantum,
  title={Quantum signatures of chaos},
  author={Haake, Fritz},
  year={1991},
  doi={10.1007/978-3-319-97580-1},
  publisher={Springer}
}

@article{RevModPhys.82.3045,
  title = {Colloquium: Topological insulators},
  author = {Hasan, M. Z. and Kane, C. L.},
  journal = {Rev. Mod. Phys.},
  volume = {82},
  issue = {4},
  pages = {3045--3067},
  numpages = {0},
  year = {2010},
  month = {Nov},
  publisher = {American Physical Society},
  doi = {10.1103/RevModPhys.82.3045},
  url = {https://link.aps.org/doi/10.1103/RevModPhys.82.3045}
}

@article{parameswaran2018many,
  title={Many-body localization, symmetry and topology},
  author={Parameswaran, SA and Vasseur, Romain},
  journal={Rep. Prog. Phys.},
  volume={81},
  number={8},
  pages={082501},
  year={2018},
  doi={10.1088/1361-6633/aac9ed},
  publisher={IOP Publishing}
}

@article{PhysRevLett.109.190601,
  title = {Coherent Forward Scattering Peak Induced by Anderson Localization},
  author = {Karpiuk, T. and Cherroret, N. and Lee, K. L. and Gr\'emaud, B. and M\"uller, C. A. and Miniatura, C.},
  journal = {Phys. Rev. Lett.},
  volume = {109},
  issue = {19},
  pages = {190601},
  numpages = {5},
  year = {2012},
  month = {Nov},
  publisher = {American Physical Society},
  doi = {10.1103/PhysRevLett.109.190601},
  url = {https://link.aps.org/doi/10.1103/PhysRevLett.109.190601}
}

@article{PhysRevA.90.063602,
  title = {Coherent forward scattering in two-dimensional disordered systems},
  author = {Ghosh, S. and Cherroret, N. and Gr\'emaud, B. and Miniatura, C. and Delande, D.},
  journal = {Phys. Rev. A},
  volume = {90},
  issue = {6},
  pages = {063602},
  numpages = {12},
  year = {2014},
  month = {Dec},
  publisher = {American Physical Society},
  doi = {10.1103/PhysRevA.90.063602},
  url = {https://link.aps.org/doi/10.1103/PhysRevA.90.063602}
}

@article{PhysRevA.95.041602,
  title = {Coherent forward scattering as a signature of Anderson metal-insulator transitions},
  author = {Ghosh, Sanjib and Miniatura, Christian and Cherroret, Nicolas and Delande, Dominique},
  journal = {Phys. Rev. A},
  volume = {95},
  issue = {4},
  pages = {041602},
  numpages = {5},
  year = {2017},
  month = {Apr},
  publisher = {American Physical Society},
  doi = {10.1103/PhysRevA.95.041602},
  url = {https://link.aps.org/doi/10.1103/PhysRevA.95.041602}
}

@article{PhysRevA.90.043605,
  title = {Dynamics of localized waves in one-dimensional random potentials: Statistical theory of the coherent forward scattering peak},
  author = {Lee, Kean Loon and Gr\'emaud, Beno\^{\i}t and Miniatura, Christian},
  journal = {Phys. Rev. A},
  volume = {90},
  issue = {4},
  pages = {043605},
  numpages = {14},
  year = {2014},
  month = {Oct},
  publisher = {American Physical Society},
  doi = {10.1103/PhysRevA.90.043605},
  url = {https://link.aps.org/doi/10.1103/PhysRevA.90.043605}
}

@article{PhysRevA.95.043626,
  title = {Coherent backscattering and forward-scattering peaks in the quantum kicked rotor},
  author = {Lemari\'e, G. and M\"uller, Cord A. and Gu\'ery-Odelin, D. and Miniatura, C.},
  journal = {Phys. Rev. A},
  volume = {95},
  issue = {4},
  pages = {043626},
  numpages = {12},
  year = {2017},
  month = {Apr},
  publisher = {American Physical Society},
  doi = {10.1103/PhysRevA.95.043626},
  url = {https://link.aps.org/doi/10.1103/PhysRevA.95.043626}
}

@article{PhysRevResearch.6.L012021,
  title = {Momentum-space signatures of the Anderson transition in a symplectic, two-dimensional, disordered ultracold gas},
  author = {Arabahmadi, Ehsan and Schumayer, Daniel and Gr\'emaud, Beno\^{\i}t and Miniatura, Christian and Hutchinson, David A. W.},
  journal = {Phys. Rev. Res.},
  volume = {6},
  issue = {1},
  pages = {L012021},
  numpages = {6},
  year = {2024},
  month = {Jan},
  publisher = {American Physical Society},
  doi = {10.1103/PhysRevResearch.6.L012021},
  url = {https://link.aps.org/doi/10.1103/PhysRevResearch.6.L012021}
}

@article{dupont_quantum_2021,
	title = {Quantum {State} {Control} of a {Bose}-{Einstein} {Condensate} in an {Optical} {Lattice}},
	volume = {2},
	issn = {2691-3399},
	url = {https://link.aps.org/doi/10.1103/PRXQuantum.2.040303},
	doi = {10.1103/PRXQuantum.2.040303},
	language = {en},
	number = {4},
	urldate = {2024-09-09},
	journal = {PRX Quantum},
	author = {Dupont, N. and Chatelain, G. and Gabardos, L. and Arnal, M. and Billy, J. and Peaudecerf, B. and Sugny, D. and Guéry-Odelin, D.},
	month = oct,
	year = {2021},
	pages = {040303},
}

@article{dupont_phase-space_2023,
	title = {Phase-space distributions of {Bose}–{Einstein} condensates in an optical lattice: optimal shaping and reconstruction},
	volume = {25},
	issn = {1367-2630},
	shorttitle = {Phase-space distributions of {Bose}–{Einstein} condensates in an optical lattice},
	url = {https://dx.doi.org/10.1088/1367-2630/acaf9a},
	doi = {10.1088/1367-2630/acaf9a},
	language = {en},
	number = {1},
	urldate = {2024-09-09},
	journal = {New Journal of Physics},
	author = {Dupont, N. and Arrouas, F. and Gabardos, L. and Ombredane, N. and Billy, J. and Peaudecerf, B. and Sugny, D. and Guéry-Odelin, D.},
	month = jan,
	year = {2023},
	pages = {013012},
}

@misc{cours_Delande,
  author    = {D. Delande},
  title     = {{Kicked rotor and Anderson localization}},
  howpublished = {Boulder School on Condensed Matter Physics},
  year      = {2013},
  note      = {Lecture I},
  url = {https://boulderschool.yale.edu/sites/default/files/files/Delande-kicked_rotor_lectures_1_and_2.pdf}
}

@ARTICLE{Chirikov,
       author = {{Chirikov}, Boris V.},
        title = "{A universal instability of many-dimensional oscillator systems}",
      journal = {Phys. Rep.},
         year = 1979,
        month = may,
       volume = {52},
       number = {5},
        pages = {263-379},
          doi = {10.1016/0370-1573(79)90023-1},
       adsurl = {https://ui.adsabs.harvard.edu/abs/1979PhR....52..263C}
}

@article{Mouchet_2001,
   title={Chaos-assisted tunneling with cold atoms},
   volume={64},
   ISSN={1095-3787},
   url={http://dx.doi.org/10.1103/PhysRevE.64.016221},
   DOI={10.1103/physreve.64.016221},
   number={1},
   pages={016221},
   journal={Phys. Rev. E},
   publisher={American Physical Society (APS)},
   author={Mouchet, A. and Miniatura, C. and Kaiser, R. and Grémaud, B. and Delande, D.},
   year={2001},
   month=jun }

@book{LesHouches1989,
  author    = {M.J. Giannoni, A. Voros and J. Zinn-Justin},
  title     = {Les Houches 1989 Session LII, Chaos and Quantum Physics},
  publisher = {North-Holland},
  year      = {1991},
  pages = {189-191},
}

@article{Marinho_2018,
   title={Spectral correlations in Anderson insulating wires},
   volume={97},
   ISSN={2469-9969},
   pages={041406(R)},
   url={http://dx.doi.org/10.1103/PhysRevB.97.041406},
   DOI={10.1103/physrevb.97.041406},
   number={4},
   journal={Phys. Rev. B},
   publisher={American Physical Society (APS)},
   author={Marinho, M. and Micklitz, T.},
   year={2018},
   month=jan }

@article{Pichard1990,
  title = {Broken symmetries and localization lengths in Anderson insulators: Theory and experiment},
  author = {Pichard, Jean-Louis and Sanquer, Marc and Slevin, Keith and Debray, Philippe},
  journal = {Phys. Rev. Lett.},
  volume = {65},
  issue = {14},
  pages = {1812--1815},
  numpages = {0},
  year = {1990},
  month = {Oct},
  publisher = {American Physical Society},
  doi = {10.1103/PhysRevLett.65.1812},
  url = {https://link.aps.org/doi/10.1103/PhysRevLett.65.1812}
}

@book{gallavottiStatisticalMechanics1999,
  title = {Statistical {{Mechanics}}},
  author = {Gallavotti, Giovanni},
  year = {1999},
  publisher = {Springer},
  address = {Berlin, Heidelberg},
  doi = {10.1007/978-3-662-03952-6},
  urldate = {2025-01-10},
  copyright = {http://www.springer.com/tdm},
  isbn = {978-3-642-08438-6 978-3-662-03952-6}
}

@article{altlandFieldTheoryQuantum1996,
  title = {Field {{Theory}} of the {{Quantum Kicked Rotor}}},
  author = {Altland, Alexander and Zirnbauer, Martin R.},
  year = {1996},
  month = nov,
  journal = {Phys. Rev. Lett.},
  volume = {77},
  number = {22},
  pages = {4536--4539},
  publisher = {American Physical Society},
  doi = {10.1103/PhysRevLett.77.4536},
  urldate = {2024-12-16}
}

@article{MicklitzPRL2014,
  title = {Strong Anderson Localization in Cold Atom Quantum Quenches},
  author = {Micklitz, T. and M\"uller, C. A. and Altland, A.},
  journal = {Phys. Rev. Lett.},
  volume = {112},
  issue = {11},
  pages = {110602},
  numpages = {5},
  year = {2014},
  month = {Mar},
  publisher = {American Physical Society},
  doi = {10.1103/PhysRevLett.112.110602},
  url = {https://link.aps.org/doi/10.1103/PhysRevLett.112.110602}
}

@article{bohigasCharacterizationChaoticQuantum1984,
  title = {Characterization of {{Chaotic Quantum Spectra}} and {{Universality}} of {{Level Fluctuation Laws}}},
  author = {Bohigas, O. and Giannoni, M. J. and Schmit, C.},
  year = {1984},
  month = jan,
  journal = {Phys. Rev. Lett.},
  volume = {52},
  number = {1},
  pages = {1--4},
  publisher = {American Physical Society},
  doi = {10.1103/PhysRevLett.52.1},
  urldate = {2024-12-23}
}

@misc{zhaoObservationQuantumThermalization2024,
  author = {Zhao, Luheng and Datla, Prithvi Raj and Tian, Weikun and Aliyu, Mohammad Mujahid and Loh, Huanqian},
  year = {2024},
  month = oct,
  number = {arXiv:2403.09517},
  eprint = {2403.09517},
  primaryclass = {quant-ph},
  publisher = {arXiv},
  doi = {10.48550/arXiv.2403.09517},
  urldate = {2025-01-17},
  archiveprefix = {arXiv}
}

@article{adlerObservationHilbertSpace2024,
  title = {Observation of {{Hilbert}} Space Fragmentation and Fractonic Excitations in {{2D}}},
  author = {Adler, Daniel and Wei, David and Will, Melissa and Srakaew, Kritsana and Agrawal, Suchita and Weckesser, Pascal and Moessner, Roderich and Pollmann, Frank and Bloch, Immanuel and Zeiher, Johannes},
  year = {2024},
  month = dec,
  journal = {Nature},
  volume = {636},
  number = {8041},
  pages = {80--85},
  publisher = {Nature Publishing Group},
  issn = {1476-4687},
  doi = {10.1038/s41586-024-08188-0},
  urldate = {2025-01-17},
  copyright = {2024 The Author(s)},
  langid = {english}
}

@article{bucaNJP2020,
  title = {Quantum synchronisation enabled by dynamical symmetries and
dissipation},
  author = {Tindall, J and Sánchez Muñoz, C and Buča, B and Jaksch, D},
  year = {2020},
  journal = {New J. Phys.},
  volume = {22},
  pages = {013026},
  publisher = {IOP Publishing},
  doi = {10.1088/1367-2630/ab60f5}
}

@article{Martinez_2023,
   title={Coherent forward scattering as a robust probe of multifractality in critical disordered media},
   volume={14},
   ISSN={2542-4653},
   doi={10.21468/scipostphys.14.3.057},
   number={3},
   journal={SciPost Physics},
   publisher={Stichting SciPost},
   author={Martinez, Maxime and Lemarié, Gabriel and Georgeot, Bertrand and Miniatura, Christian and Giraud, Olivier},
   year={2023},
   month=mar }

@article{hoQuantumClassicalFloquet2023,
  title = {Quantum and Classical {{Floquet}} Prethermalization},
  author = {Ho, Wen Wei and Mori, Takashi and Abanin, Dmitry A. and Dalla Torre, Emanuele G.},
  year = {2023},
  month = jul,
  journal = {Annals of Physics},
  volume = {454},
  pages = {169297},
  issn = {0003-4916},
  doi = {10.1016/j.aop.2023.169297},
  urldate = {2025-01-23},
}

@article{mooreObservationDynamicalLocalization1994e,
  title = {Observation of {{Dynamical Localization}} in {{Atomic Momentum Transfer}}: {{A New Testing Ground}} for {{Quantum Chaos}}},
  shorttitle = {Observation of {{Dynamical Localization}} in {{Atomic Momentum Transfer}}},
  author = {Moore, F. L. and Robinson, J. C. and Bharucha, C. and Williams, P. E. and Raizen, M. G.},
  year = {1994},
  month = nov,
  journal = {Phys. Rev. Lett.},
  volume = {73},
  number = {22},
  pages = {2974--2977},
  publisher = {American Physical Society},
  doi = {10.1103/PhysRevLett.73.2974},
  urldate = {2025-01-23},
}

@article{SHEPELYANSKY1987103,
title = {Localization of diffusive excitation in multi-level systems},
journal = {Physica D: Nonlinear Phenomena},
volume = {28},
number = {1},
pages = {103-114},
year = {1987},
issn = {0167-2789},
doi = {https://doi.org/10.1016/0167-2789(87)90123-0},
url = {https://www.sciencedirect.com/science/article/pii/0167278987901230},
author = {D.L. Shepelyansky}
}

@article{PhysRevLett.44.1586,
  title = {Calculation of Turbulent Diffusion for the Chirikov-Taylor Model},
  author = {Rechester, A. B. and White, R. B.},
  journal = {Phys. Rev. Lett.},
  volume = {44},
  issue = {24},
  pages = {1586--1589},
  numpages = {0},
  year = {1980},
  month = {Jun},
  publisher = {American Physical Society},
  doi = {10.1103/PhysRevLett.44.1586},
  url = {https://link.aps.org/doi/10.1103/PhysRevLett.44.1586}
}

@article{PhysRevLett.69.217,
  title = {Symmetry breaking and localization in quantum chaotic systems},
  author = {Bl\"umel, R. and Smilansky, U.},
  journal = {Phys. Rev. Lett.},
  volume = {69},
  issue = {2},
  pages = {217--220},
  numpages = {0},
  year = {1992},
  month = {Jul},
  publisher = {American Physical Society},
  doi = {10.1103/PhysRevLett.69.217},
  url = {https://link.aps.org/doi/10.1103/PhysRevLett.69.217}
}

@article{Prigodin1994,
  title = {Mesoscopic dynamical echo in quantum dots},
  author = {Prigodin, V. N. and Altshuler, B. L. and Efetov, K. B. and Iida, S.},
  journal = {Phys. Rev. Lett.},
  volume = {72},
  issue = {4},
  pages = {546--549},
  year = {1994},
  doi = {10.1103/PhysRevLett.72.546},
  url = {https://link.aps.org/doi/10.1103/PhysRevLett.72.546}
}

@article{Weaver2000,
  title = {Enhanced Backscattering and Modal Echo of Reverberant Elastic Waves},
  author = {Weaver, Richard L. and Lobkis, Oleg I.},
  journal = {Phys. Rev. Lett.},
  volume = {84},
  issue = {21},
  pages = {4942--4945},
  numpages = {0},
  year = {2000},
  month = {May},
  publisher = {American Physical Society},
  doi = {10.1103/PhysRevLett.84.4942},
  url = {https://link.aps.org/doi/10.1103/PhysRevLett.84.4942}
}

@misc{Hebraud25,
author={{Hebraud et al.}, J.},
note={(in preparation)}
}

\begin{appendix}

\renewcommand{\theequation}{A\arabic{equation}}
\renewcommand{\thetable}{A\arabic{table}}
\renewcommand{\thefigure}{A\arabic{figure}}
\setcounter{equation}{0}
\setcounter{table}{0}
\setcounter{figure}{0}

\section*{Methods}

\subsection*{Experimental setup}

Our experimental setup produces Bose-Einstein condensates (BEC) of rubidium-87 in a hybrid (magnetic and dipolar) trap, with weak harmonic trapping (angular frequencies ($\Omega_x,\Omega_y, \Omega_z$)=$2\pi\times({7,73,66})\,{\rm Hz}$)~\cite{dupont_quantum_2021}. This confinement does not affect the dynamics over the timescale of the experiments. A 1D optical lattice with spatial period $d=532$ nm is superimposed to the hybrid trap on the $x$-axis. The optical lattice is produced by the interference of two far-detuned counter-propagating beams derived from the same laser. An acousto-optic modulator (AOM) controls the laser amplitude, while two other phase-locked AOMs placed on each lattice beam control their relative phase. The lattice depth $s$ and phase $\varphi$ can thus be arbitrarily and independently time-modulated. The bandwidth for amplitude or phase modulation is about $1\,\mathrm{MHz}$. The Hamiltonian describing the dynamics in the lattice writes

\begin{equation}
	\hat{H} = \frac{\hat{P}^2}{2m} - \frac{s E_L}{2} \cos\left( k_L \hat{X} + \varphi(T) \right)f(T),
\end{equation}
with $m$ the atomic mass, $k_L=2\pi/d$ the lattice wavevector and $E_L=\hbar^2 k_L^2/2m=h\nu_L$ the lattice characteristic energy, with the characteristic lattice frequency $\nu_L$. The maximum reachable lattice depth is $s\simeq 40$.

\subsection*{System modeling} 

To model the experimental wavefunction, which occupies a finite number of lattice sites, it is written as a superposition
\begin{equation}
	\Psi(x,t)=\int_{-0.5}^{0.5}\mathrm{d}\beta \Phi(\beta) \Psi_{\beta}(x,t),
\end{equation}
where $\abs{\Phi(\beta)}^2$ is a narrow quasi-momentum distribution centered on $\beta=0$ with $\int_{-0.5}^{0.5}\abs{\Phi(\beta)}^2\mathrm{d}\beta=1$, and the components $\Psi_{\beta}(x,t)$ are normalized wavefunctions evolving in the subspace of quasi-momentum $\beta$, which can be decomposed on a set of plane waves:
\begin{equation}
	\Psi_{\beta}(x,t)=\sum_\ell c_{\ell,\beta}(t)\frac{1}{\sqrt{2\pi}}e^{i(\ell+\beta) x}.
\end{equation}
The planewave $\ket{\ell+\beta}$ is an eigenstate of momentum with eigenvalue $p=\left( \ell+\beta\right) \hbar_\mathrm{eff}$ and wavefunction $\braket{x}{\ell+\beta}=\frac{1}{\sqrt{2\pi}}e^{i(\ell+\beta) x}$.

Experimentally, we measure the momentum population after a time-of-flight. This measurement does not resolve quasi-momentum components and provides the momentum density averaged over quasi-momenta near integer multiples of $\hbar_\mathrm{eff}$, $P(p\simeq\ell\hbar_\mathrm{eff},t)=\int\mathrm{d}\beta |\Phi(\beta)|^2  |c_{\ell,\beta}(t)|^2$.
Since we measure an average of all lattice sites contributions, the corresponding spatial density is the one-site density averaged over all sites, $P(x,t)$.

The average across multiple experiments is done by taking a statistical average for both densities over the evolution obtained from several modulation functions.

Modeling $\abs{\Phi(\beta)}^2$ as a rectangular distribution centered at $\beta_m=0$ with a width of $\Delta\beta=0.1$ gives good agreement with experiments. This distribution is discretized over a number of quasi-momenta, $N_\beta$, which depends on the simulation and is chosen to ensure that no boundary effects arise from the dynamics in position space. Typically $N_\beta=51$.

Simulations use a discretization of the evolution operator from the time-dependent Schrödinger equation (TDSE), $\hat{U}(t,0)=\hat{U}(t,t-\delta t)...\hat{U}(\delta t,0)$. The modulation is considered piecewise-constant, with a fixed value during a time step $\delta t$.
The typical number of time-steps is $500$ per period. The size of the Hilbert space chosen for computation is also adapted to the dynamics to avoid any boundary effects.

Over the typical duration of the experimental sequence (between 1 and 2 ms), the BEC does not experience any noticeable decoherence, hence our model does not require the introduction of any phenomenological decay time, in contrast to previous kicked rotor experiments~\cite{hainaut2018controlling}.

\subsection*{Characteristic lengths: $L$ and $\xi_{\mathrm{loc}}$}

In a 1D disordered system, localization of a wave-packet occurs and results in a freezing of the diffusion. In our system this diffusion takes place in momentum space where the localized average density can be approximated as $P(p)=Ae^{-|p|/\xi_{\mathrm{p}}}$.  Two characteristic lengths can be measured : the localization length $\xi_{p}$, defined as the scale of the exponential decay of the wavefunction in the absence of a boundary, and the momentum extension of the chaotic sea $L_p$, defining the extent in momentum of the area accessible through diffusion. These values can also be expressed in $\hbar_\mathrm{eff}$ units : $L=L_p/\hbar_\mathrm{eff}$ and $\xi_{loc}=\xi_p/\hbar_\mathrm{eff}$.

\begin{figure}[ht!]
	\centering
	\includegraphics[width=\linewidth]{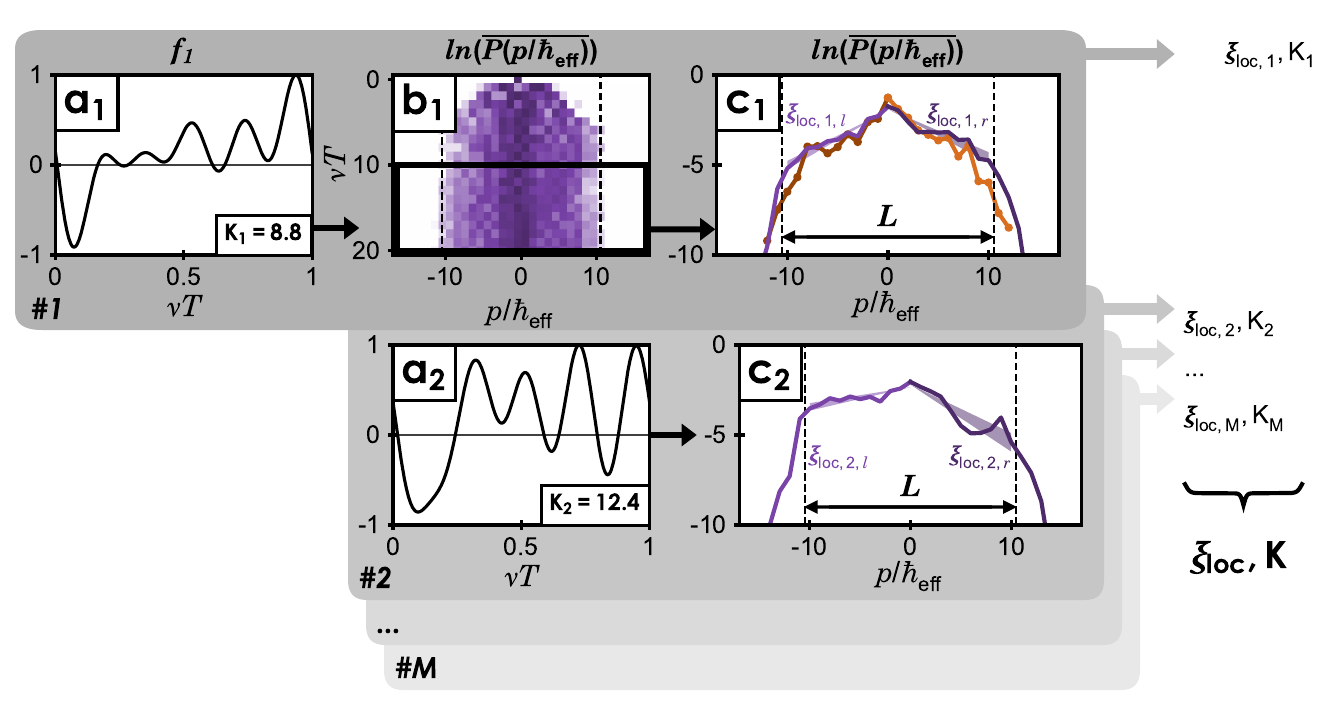}
	\caption{\textbf{Estimation of the mean localization length $\xi_{\mathrm{loc}}$ and stochastic coefficient $K$.} First row: $\mathbf{a_1}$ shows an example of modulation function and its stochastic coefficient $K_{1}$, and $\mathbf{b_1}$ the computed numerical evolution of the momentum density starting from the initial zero-momentum state (the condensate). $\mathbf{c_1}$ shows the time-averaged momentum density from period 10 to 20 obtained both numerically (purple) and experimentally (orange), showing good agreement. The slopes and sharp decrease indicate the localization length and classical box length $L$ respectively. Linear fits on the two slopes give left $\xi_{\mathrm{loc},1,l}$ and right $\xi_{\mathrm{loc},1,r}$ localization lengths. 
    Second row: Similar computations can be performed for each modulation function $f_m$. Taking the mean values give the average $\xi_{\mathrm{loc}}$ and $K$ for this parameter regime. Parameter values are the same as for Figure \ref{fig:Fig2_measurement} of the main text (see table \ref{tab:experimentalpara}), in a broken symmetry case ($A_\varphi=2$). Parameters: $L=20.22 \pm 1.54 $, $\xi_{\rm loc}=6.53\pm 2.87$.
	}
	\label{fig:Fig5_LongueurLoc}
\end{figure}

These parameters are determined numerically. The chaotic dynamics starting from an initial momentum state $\ket{p=0}$ is simulated, with a lattice modulated by a function $f_m(t)$ ($1\leq m \leq M$) and a stochastic parameter $K_{m}>9$. 
The localization length $\xi_{\mathrm{loc}, m}$ is estimated by fitting and taking the mean value from the right and left slopes of the average logarithmic momentum density $\ln(P(p))$ obtained from multiple periods once localization has set in. In most of the results, the average is taken over $70$ periods and starts after period $30$.

The measurement method used in this paper requires multiple modulation functions, each one having a different stochastic coefficient and leading to a different localization length. The estimation of an overall averaged localization length $\xi_{\mathrm{loc}}$ is given by taking the mean value of all $\xi_{\mathrm{loc}, m}$. We also define a mean stochastic coefficient $K$.

The momentum extension of the chaotic sea, $L$, is also deduced from the same computation. It is defined as the interval between the left and right sharp drops of the momentum log-density average $\overline{\ln(P(p))}$ over all modulation functions. The values obtained correspond well to the extent of the chaotic sea of the classical phase space.

An example of the determination of $\xi_{\mathrm{loc}}$ and $L$ is given in Extended data Figure~\ref{fig:Fig5_LongueurLoc}, and all experimental parameters corresponding to the characteristic lengths $L$ and $\xi_{\mathrm{loc}}$ mentioned in the Figures of the main text are referenced in Extended data Table \ref{tab:experimentalpara} along with the average stochastic parameter $K$.

\begin{table*}[ht!]
	\centering
	\begin{tblr}{hline{1,2,Z} = {1pt},hline{4,5,V} = {0.5pt,dashed},vline{1,Z} = {1pt}, colspec={ccccccc}, rowspec={c}}
		&  $\mathit{s}$&  $\mathbf{\nu}$& $\mathbf{N_H}$& $\mathbf{A_{\varphi}}$ &$\mathbf{K}$ & $\hbar_\mathrm{eff}$\\
            \textbf{Fig.1} $\mathbf{b}$&  $25.8\pm0.22$&  $10.729\,$kHz& 4& 1 & $110\pm7$& $9.5$ \\
		\textbf{Fig.1} $\mathbf{c}$&  $26.3\pm0.06$&  $10.673\,$kHz& 4& - & $113\pm7$ & $9.55$ \\
        \textbf{Fig.2} &  $24.2\pm0.28$&  $35.050\,$kHz& 5& - & $8.59\pm0.73$ & $2.91$ \\
		\textbf{Fig.3} $\mathbf{a_2}$&  $27.4\pm0.28$&  $10.508\,$kHz& 4&2& $125\pm12$& $9.7$ \\
        \textbf{Fig.3} $\mathbf{b_2}$&  $ 26.1\pm0.63$&  $10.729\,$kHz& 4&1& $110\pm7$& $9.5$  \\
		\textbf{Fig.3} $\mathbf{c_2}$&  $27.9\pm0.38$&  $10.454\,$kHz& 4& 2 & $122\pm8$ & $9.75$ \\
		\textbf{Fig.3} $\mathbf{d_2}$&   $26.5\pm0.1$&  $10.673\,$kHz& 4& - & $113\pm7$ & $9.55$ \\
		\textbf{Fig.3} $\mathbf{a_3}$&  $28.7\pm0.45$&  $12.741\,$kHz& 27&2& $20.3\pm1.4$& $8$ \\
         \textbf{Fig.3} $\mathbf{b_3}$&  $6.3\pm0.13$&  $12.430\,$kHz& 15&$1.5$& $10.9\pm0.2$& $8.2$ \\
    $\,$ &  $s_{rot}=26.8\pm0.3$&  $\,$& $\,$&$\,$& $\,$ \\
		\textbf{Fig.3} $\mathbf{c_3}$&  $23.9\pm0.25$&  $26.823\,$kHz& 17&2& $9.0\pm0.1$& $3.8$ \\
		
		\textbf{Fig.3} $\mathbf{d_3}$&  $5.2\pm0.14$&  $10.618\,$kHz& 20& - & $11.2\pm0.1$& $9.6$ \\
		$\,$ &  $s_{rot}=22.3\pm0.4$&  $\,$& $\,$&$\,$& $\,$ \\
		\textbf{Fig.4} $\mathbf{a}$& $11.3\pm0.18$& $7.078\,$kHz&15&4& $43.3\pm5.4$& $14.4$ \\
		\textbf{Fig.4} $\mathbf{b}$& $26.3\pm0.29$& $5.926\,$kHz&13&2& $171\pm15$& $17.2$ \\
		\textbf{Fig.4} $\mathbf{c}$& $33.9\pm0.32$& $10.193\,$kHz&9&2&$101\pm10$& $10$ \\
        \textbf{Fig.4} $\mathbf{d}$& $29.67\pm0.17$& $10.193\,$kHz&9&2&$103\pm1$& $10$ \\
	\end{tblr}
	\caption{\textbf{Experimental parameters corresponding to the results presented in the figures of the main text.} The error bars represent one standard deviation across all averaged experiments. $A_\varphi$ is the amplitude of additional lattice phase modulation $\varphi(t)$. Specific values of the lattice depth for the phase space rotation $s_{rot}$ are indicated, when it differs from $s$.}
	\label{tab:experimentalpara}
\end{table*}

\subsection*{Initial state preparation using Optimal Control}
\label{subsec:OptimalControlPreparation}

Measurements of the CBS and CFS peaks in the shaken rotor require a peaked initial state at $x \neq 0$. The phase space rotation measurement also requires a finite momentum extension, contained within the closed trajectories of the static sine-potential phase space. 
Such a state can be defined using the lattice squeezed Gaussian state
\begin{equation}
	\ket{\tilde{G}_{\beta}(x_c,p_c,s,\sigma)}=\sum_{\ell\in \mathbb{Z}}c_{\ell,\beta}(x_c,p_c,s,\sigma)\ket{\ell+\beta},
\end{equation}
with coefficients
\begin{equation}
	c_{\ell,\beta}(x_c,p_c,s,\sigma)=\frac{2\sigma^2}{\pi \sqrt{s}} e^{ix_c p_c/2}e^{-i(\ell+\beta) x_c}e^{-\sigma^2(\ell+\beta-p_c)^2/\sqrt{s}},
\end{equation}
where $\beta$ is the quasi-momentum, $x_c$ and $p_c$ the Gaussian state average position and momentum in a lattice well, $s$ the lattice depth, and $\sigma$ the squeezing factor quantifying the Gaussian spatial extension $\Delta x = \sigma \Delta x_0$ with $\Delta x_0=k_L^{-1}s^{-1/4}$ the non-squeezed extension, similar to the extension of the ground-state at depth $s$. 

We choose $\ket{\Psi_{i}}=\ket{\tilde{G}_0(-\pi/2,0,s,0.6)}$ as our initial state, with $s$ the depth used during the phase space rotation. It corresponds to an extension in dimensionless position units of $\Delta x = 0.27$ for $s=25$, corresponding to the ground-state width in a lattice with effective depth $s_\mathrm{eff}\sim193$~\cite{dupont_phase-space_2023}.

This state is prepared through optimal control: starting from an state $\ket{\Psi_0}$, a desired target state $\ket{\Psi_{T}}$ can be reached using a lattice phase modulation $\varphi_{\mathrm{oc}}(t)$ determined by a quantum Optimal Control (OC) algorithm \cite{dupont_quantum_2021}, which uses gradient ascent to maximize the fidelity to the target $F=|\braket{\Psi_0}{\Psi_T}|^2$.
For all our experiments, $\ket{\Psi_0}$ is the ground-state of the lattice at depth $s_\mathrm{oc}=5$, and the target is a squeezed Gaussian state $\ket{\Psi_{T}}=\ket{\tilde{G}_0(0,0,s,0.6)}$ centered in phase space. The OC preparation is determined at depth $s_\mathrm{oc}$ for a duration close to $100\,\mu s$
and numerical fidelity $F\geq0.99$ in subspace $\beta=0$.
After the preparation, the lattice phase is shifted with $\pi/2$ to place the Gaussian at $x=-x_0=-\pi/2$ to reach the desired initial state $\ket{\Psi_i}$. 

\subsection*{Quantum state reconstruction}

The final experimental state can be characterized using a maximum likelihood reconstruction algorithm to determine its density matrix \cite{dupont_phase-space_2023}. The algorithm uses dynamics of the state in a static lattice to iteratively transform an initial-guess density matrix $\hat{\rho}_0$ until it converges to the most likely one, $\hat{\rho}_{ML}$.

The likelihood function is defined with respect to the system's density matrix as:  
\begin{equation}
	\mathcal{L}\left[\hat{\rho}\right] = \prod_{\ell,\tau} p_\ell(\tau)^{f_\ell(\tau)},
\end{equation}  
with $p_\ell(\tau)$ the expected populations of momentum $\ell$ at time $\tau$, as obtained from $\hat{\rho}$, raised to the power of \( f_\ell(\tau) \), the corresponding experimentally measured populations. The likelihood reaches its maximum when the probabilities from $\hat{\rho}$ match the measurements.

The maximum likelihood algorithm
introduces the operator in the zero quasi-momentum subspace  
\begin{equation}
	\hat{R}\left[\hat{\rho}\right] = \sum_{\ell,\tau} \frac{f_\ell(\tau)}{p_\ell(\tau)} \hat{\varepsilon}_\ell(\tau),
\end{equation}  
with $\hat{\varepsilon}_\ell(\tau)=(1/N_\tau)\hat{U}^\dagger(\tau,0) \ket{\ell}\bra{\ell}\hat{U}(\tau,0)$ forming a positive operator-valued measure (POVM). Repeated application of $\hat{R}$ to the initial guess iteratively increase the likelihood, until a fixed point  $\hat{\rho}_{ML}$ is reached, that maximizes the likelihood.

\bigskip

\begin{figure*}[ht!]
	\centering
	\includegraphics[width=0.9\textwidth]{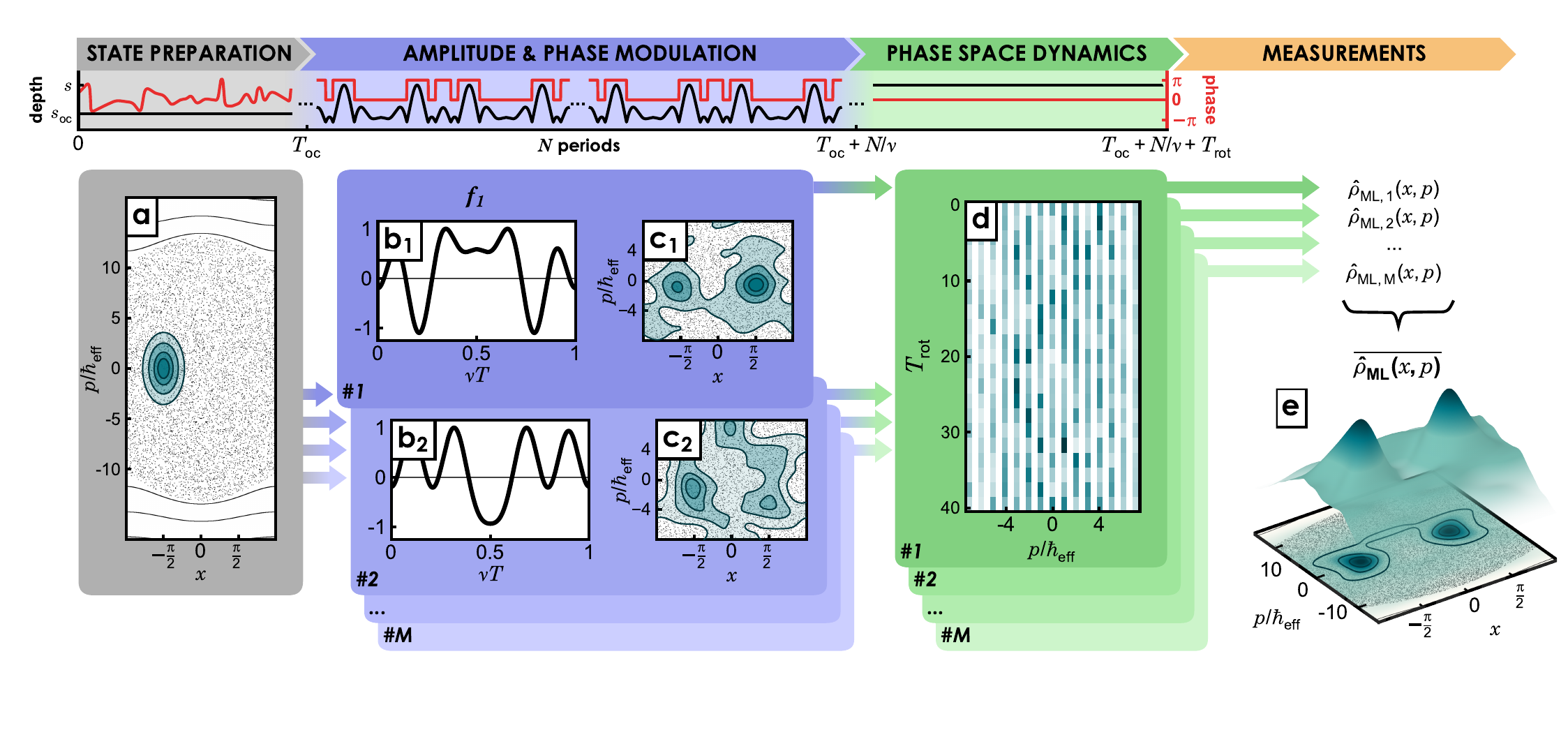}
	\caption{\textbf{Adapted experimental protocol for reconstruction.} Numerical results adapted from Figure \ref{fig:Fig2_measurement}. The preparation and modulation remain unchanged. After modulation, the state evolution is probed in a static lattice of depth $s$ (with no phase shift, $\varphi=0$) over a duration of $40\,\mu$s with steps of $2\,\mu$s. The associated density matrix $\hat{\mathit{\rho}}_{ML,m}$ are reconstructed using a maximum likelihood algorithm~\cite{dupont_phase-space_2023}. The final density matrix is obtained by averaging over all modulation functions. The average Husimi distribution $\overline{\hat{\mathit{\rho}}_{ML}}$ shown in $\mathbf{e}$ exhibits both CBS and CFS peaks.}
	\label{fig:MeasurementMethod_Reconstruction}
\end{figure*}

To achieve this experimentally, the procedure is almost the same as the phase space method presented in the main text. After the chaotic dynamics, all modulations are turned off, and the state dynamics is probed in a static lattice ($\varphi=0$ at given depth $s$) with a constant time-step of $2\,\mathrm{\mu s}$ for a total of $N_t=21$ steps. This process is depicted in Extended data Figure \ref{fig:MeasurementMethod_Reconstruction}.

However, since our system has a finite quasi-momentum width, the measurements correspond to averaged populations. 
For chaotic dynamics each quasi-momentum subspace has a significantly different population evolution which cannot be resolved in our measurements: the density matrix obtained from the reconstruction, which only accounts for $\beta=0$ while using the measured average momentum populations, will have a limited purity $\gamma=\tr\left\lbrace \hat{\rho}^2\right\rbrace<1$. 
The typical purity for an individual state reconstruction contributing to Figure \ref{fig:Fig1_schematicillustration} $ {\bf b}$ and ${\bf c}$ is $\gamma\simeq0.26$. These density matrix representations provide a good qualitative assessment, such as identifying the presence or absence of peaks, but do not allow for a quantitative estimation.

The initial state, prepared from the lattice ground state with the optimal control procedure,
is less sensitive to the quasi-momentum width, enabling reliable reconstruction in an effective $\beta=0$ subspace~\cite{dupont_phase-space_2023}. For the initial state, we get a reconstructed purity $\gamma=0.93$ and fidelity $\mathcal{F}_{\mathrm{rec}}=\bra{\Psi_T}\hat{\rho_{ML}}\ket{\Psi_T}=0.91$ to the desired target.

\subsection*{Averaged Husimi distribution}

Once the density matrix is determined, its Husimi representation is deduced using
\begin{equation}
	H_{\hat{\rho}}(x,p) = \frac{1}{2\pi} \langle \tilde{G}_0(x,p,s,1) |\hat{\rho}_{ML} |\tilde{G}_0(x,p,s,1)\rangle
\end{equation}
which is a projection on a periodic Gaussian state with $\beta=0$ centered on $x$ and $p$.

In this work, the interesting results emerge from averaged signals over different modulation functions. The corresponding Husimi distribution is experimentally obtained by repeating the reconstruction process for each different modulation function, and averaging all associated Husimi distributions. The final mean Husimi can be represented on the phase space of a lattice cell, and reflects the average distribution over the lattice.

In the numerical modeling, the computed Husimi distribution corresponds to the average distribution over all lattice sites, and is found by averaging the Husimi distributions obtained from each considered quasi-momentum component together.

\subsection*{Phase space rotation}

\begin{figure*}[ht!]
	\centering
	\includegraphics[width=0.75\textwidth]{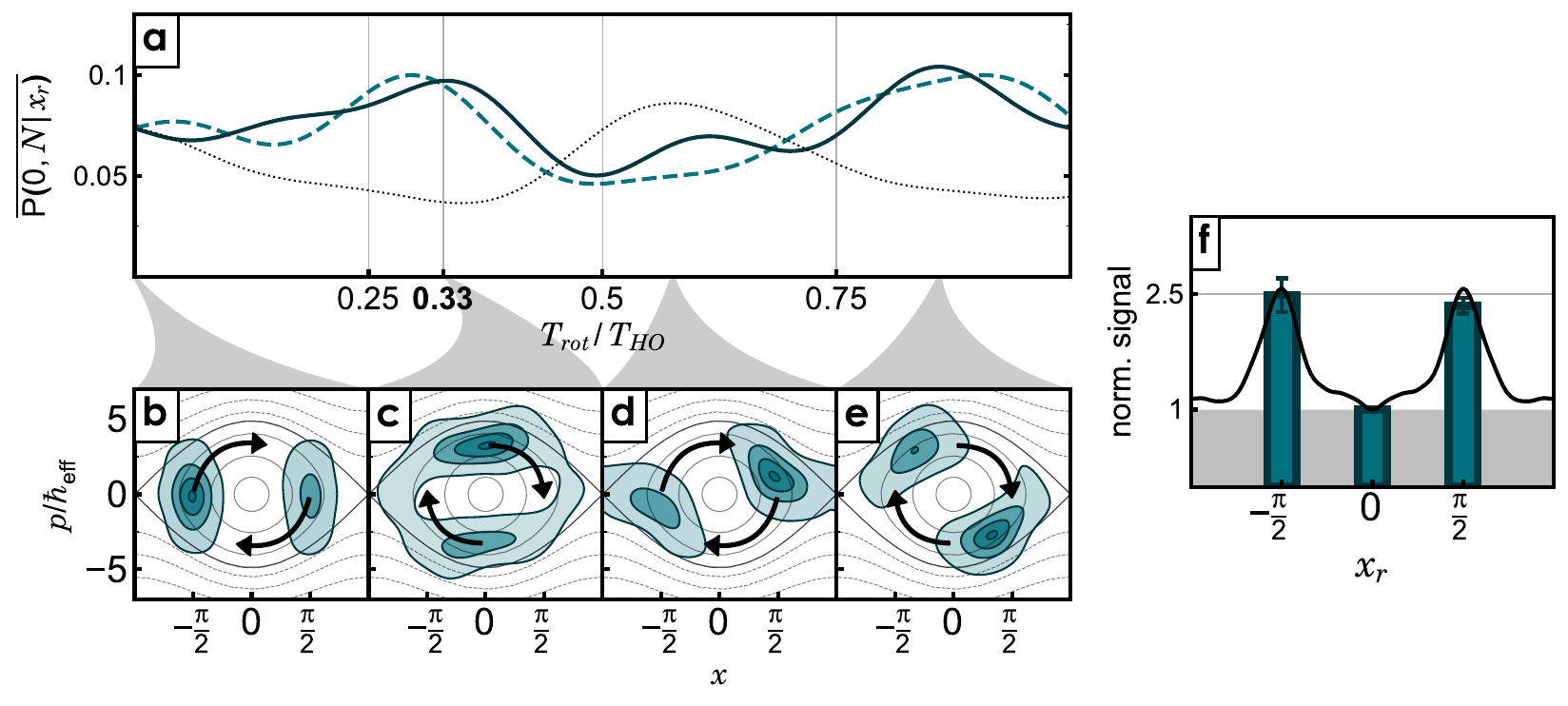}
	\caption{\textbf{Phase space rotation measurement}. After the modulation stage, the state is held in a static lattice and undergoes a phase space rotation. \textbf{(a)} The post-rotation signal, defined as the measured average population in momentum 0, is shown for rotation centers $x_r=-\pi/2$ (continous line), $x_r=\pi/2$ (dashed line) and $x_r=0$ (dotted line). \textbf{(b-e)}  Husimi distribution at different rotation times for $x_r=0$ : at $33\%$ of the harmonic oscillator period $T_{HO}$, the two-peaked spatial distribution is mapped into the momentum one. The optimal rotation time is determined as the shortest one maximizing the averaged population in momentum 0 for CBS and/or CFS. \textbf{(f)} A normalized post-rotation signal is defined as the measured average population in momentum 0 for $x_r=\pm\pi/2$ (associated to CFS or CBS) over that for $x_r=0$. Error bars correspond to one standard deviation of the mean.}
	\label{fig:RotationTime}
\end{figure*}

The phase space rotation measurement relies on a harmonic approximation of a lattice well at large depth $s$ and consists of three phase space rotations of angle $\simeq\pi/2$, centered at positions $x_r=-x_0=-\pi/2$, $x_r=x_0$ or intermediate position $x_r=0$, to respectively probe CFS, CBS and background.

The evolution of a state peaked at the center of the harmonic oscillator potential during a quarter of its period $T_{HO}=1/(\sqrt{s}\nu_L)$ converts the position distribution into the momentum distribution. In the finite-depth, anharmonic lattice well, this mapping requires a rotation duration which is not exactly equal to a quarter of $T_{HO}$, and depends on the depth $s$ and the considered state.

The optimized rotation time is numerically determined for each measurement. This optimal time corresponds to the one that maximizes the population $\overline{P(p = 0)}$ obtained for $x_r = -\pi/2$, averaged over three consecutive modulation periods and over the modulation functions. When CBS is present in the system, the rotation time is taken as the mean between the optimal ones for CFS and CBS. An illustration of this optimization is provided in Extended data Figure \ref{fig:RotationTime}.
The obtained signal reflects well the expected peak heights at infinite time in position space~\cite{PhysRevA.95.043626} (see Fig. \ref{fig:RotationTime} \textbf{f}).

Experimentally, to ensure the reproducibility of our data against experimental fluctuations, the populations corresponding to each rotation are measured twice. The absolute difference between the distributions of a pair is computed and compared to a numerical threshold, above which the pair is discarded. This leads on average to the conservation of between 75\% and 100\% of the data for each parameter set.

\subsection*{Studied symmetries}

The Hamiltonian from equation (1) in the main text is an adapted kicked rotor model, in which we have additional control using the amplitude modulation function $F(t)$ and the lattice phase $\varphi(t)$. Both are used to control our system symmetries. 

\begin{table*}[ht!]
	\centering
	\begin{tblr}{hline{1,2,Z} = {1pt},hline{X} = {0.5pt},vline{1,Z} = {1pt}, colspec={cccc}, rowspec={c}}
		
		\textbf{None}& $\mathsf{PT}$& $\mathsf{T}$&  $\mathsf{P+T}$   \\
		$t\rightarrow t$&  $t\rightarrow -t$&  $t\rightarrow -t$& $t\rightarrow -t$\\
		$p\rightarrow p$&  $p\rightarrow p$&  $p\rightarrow -p$& $p\rightarrow -p$\\
		$x\rightarrow x$&  $x\rightarrow -x$&  $x\rightarrow x$& $x\rightarrow -x$\\
		$F(t)\neq F(1-t)$&  $F(t)= F(1-t)$&  $F(t)= F(1-t)$& $F(t)= F(1-t)$\\
		$\varphi(t)=-\varphi(-t)$&  $\varphi(t)=-\varphi(-t)$&  $\varphi(t)=\varphi(-t)$& $\varphi(t)=0$\\
	\end{tblr}
	\caption{\textbf{All studied symmetries.} The first row lists the symmetry names, the second row their corresponding $x$, $p$ and $t$ transformations, and the last row details the corresponding choice for time-reversibility of the modulation function $F(t)$ and the parity of additional lattice phase modulation $\varphi(t)$. }
	\label{tab:symmetries}
\end{table*}

The time-reversal symmetry ($\mathsf{T}$) is easily broken by choosing the harmonics phases $\phi_n \in \left[0,\pi\right] $ with at least one of them not zero or $\pi$, implying that $F(t)\neq F(1-t)$. The $\mathsf{T}$-symmetry holds when $F(t)= F(1-t)$, meaning that all $\phi_n \in \left\lbrace 0,\pi\right\rbrace$.\\
The parity symmetry ($\mathsf{P}$) is broken when a lattice phase modulation $\varphi(t)$ is used. To reach a desired symmetry, the phase modulation function can be chosen even, 	
$\varphi(t)=A_\varphi\left(1-\cos(2\pi t)\right)/2$, or odd $\varphi(t) =  2A_\varphi/(3\sqrt{3}) \left(\sin(2\pi t) - \sin(4\pi t)/2\right)$, with $A_\varphi$ a phase modulation amplitude. These phase modulations are chosen to contain a minimum number of harmonics while ensuring a null derivative at $t=0$ in order to guarantee a continuous variation of momentum in the lattice.
All tested symmetries are referenced in Extended data Table \ref{tab:symmetries}.

\subsection*{Comparison between the shaken rotor and the kicked rotor}
\label{sup:theory}

In this section, we numerically study the properties of classical chaotic diffusion and quantum dynamical localization in the kicked rotor and the shaken rotor. 
We consider two shaken rotor models: one without symmetry, where the modulation functions are asymmetric in time, and one with \(\mathsf{PT}\) symmetry, where the amplitude modulation is time-symmetric and the phase modulation is time-antisymmetric (see Extended Data Table E2).

\subsubsection{Classical dynamics}
In the kicked rotor, for sufficiently large kicking strength \( K \gtrsim 0.97 \), classical chaotic trajectories are not confined by regular structures in phase space~\cite{Chirikov} and ergodically explore the entire chaotic region. This results, on average, in diffusive transport in momentum space~\cite{cours_Delande, ott2002chaos}:  
\begin{equation}
\langle p^2(t) \rangle \approx 2D_{cl} t, \label{eq:classdiff} 
\end{equation}
where the average $\langle ...\rangle$ is taken over initial conditions with \( p=0 \) and \( x \) randomly sampled in \( [0,2\pi] \).  

For the shaken rotor, the smooth modulation introduces a crucial difference. The Dirac comb of the kicked rotor's temporal forcing is replaced by a modulation function \( F(t) \) containing only a finite number of frequencies \( N_H \).  
As a result, regular trajectories enclose a chaotic sea of finite extent in momentum space, approximately \( L_p \approx 2\pi(2N_H+1) \) (with an actual size also dependent on \( K \)), see e.g.~\cite{Mouchet_2001}. Consequently, in the shaken rotor, the classical diffusive transport described by Eq.~\eqref{eq:classdiff} persists only for a finite duration before \( \langle p^2(t) \rangle \) saturates at a finite value.  

\begin{figure}[t]
	\centering
	\includegraphics[width=\linewidth]{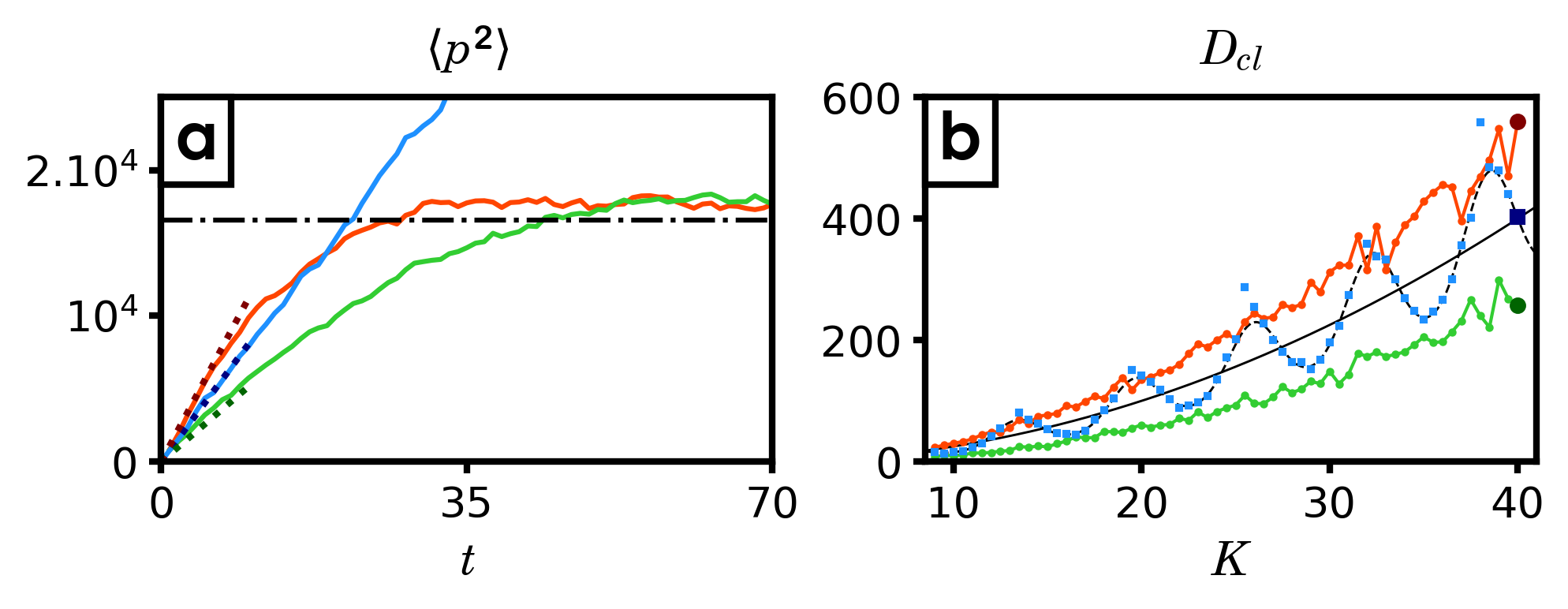}
	\caption{\textbf{Classical dynamics in the kicked rotor and shaken rotor models.}  
\textbf{(a)} The classical diffusive transport in momentum space for the kicked rotor with large kicking strength \( K = 40 \) follows Eq.~\eqref{eq:classdiff}, as shown by the blue curve. In the shaken rotor with \( K = 40 \) and \( N_H = 35 \), corresponding to a chaotic sea of finite size \( L_p \approx 446 \), diffusion saturates at large times to \( \langle p^2 \rangle_{\text{lim}} \approx L_p^2/12 \) (dashed-dotted horizontal line, see text). The case in the \( \mathsf{PT} \)-symmetry class is shown in green, while the case in the \( \varnothing \)-symmetry class is in red. Data are averaged over 1000 initial conditions \( \{ x_0, p_0 = 0 \} \), with \( x_0 \) uniformly distributed in \([0, 2\pi]\), and for the shaken rotor, over 100 different modulation functions. Dotted colored lines indicate linear fits at short times (\( t<10 \)) using Eq.~\eqref{eq:classdiff} to extract the classical diffusion coefficient \( D_{cl} \), whose values are shown in panel \textbf{(b)}.  
\textbf{(b)} Classical diffusion coefficient \( D_{cl} \) versus \( K \) in the three models. For the kicked rotor (blue symbols), \( D_{cl} \) follows the analytical expression  
$
D_{cl} \approx \frac{K^2}{4} \left( 1 - 2J_2(K) + 2J_2(K)^2 \right)
$
(black dashed line, see ~\cite{PhysRevLett.44.1586}), where \( J_2 \) is the ordinary Bessel function of the second order. In the shaken rotor, averaging over different modulation functions suppresses this oscillatory behavior, leading to a scaling \( D_{cl} \propto K^2 \). The black line represents \( D_{cl} = K^2/4 \) for both the shaken rotor in the \( \mathsf{PT} \)-symmetry class (green symbols) and in the \( \varnothing \)-symmetry class (red symbols).
	}
	\label{fig:CompareKR_Classical}
\end{figure}

This is illustrated in Fig.~\ref{fig:CompareKR_Classical} \textbf{a}, where the kicked rotor with \( K=40 \) exhibits unbounded diffusive transport in momentum space, while for the shaken rotor, \( \langle p^2(t) \rangle \) saturates at long times to \( \langle p^2 \rangle_{\text{lim}} \approx L_p^2/12 \), corresponding to a uniform distribution over the finite chaotic sea.  
The classical diffusion coefficient \( D_{cl} \), extracted at short times by fitting Eq.~\eqref{eq:classdiff} to numerical data, exhibits similar dependence on \( K \) in both models, as shown in Fig.~\ref{fig:CompareKR_Classical} \textbf{b}.

\subsubsection{Quantum dynamical localization} 
In the quantum regime, the evolution of a wave packet initially peaked at \( p=0 \) resembles classical diffusion only at short times. At longer times, dynamical localization in momentum space sets in~\cite{Casati79, mooreObservationDynamicalLocalization1994e}: the wave packet stabilizes into a stationary, exponentially localized distribution characterized by a localization length \( \xi_p \) in momentum space. Consequently, the variance of the wave packet saturates at \( \langle p^2 \rangle \approx  2 \xi_p^2 \). The same phenomenon occurs in the shaken rotor in the limit \( L_p \gg \xi_p \).  

\begin{figure}[h]
	\centering
	\includegraphics[width=\linewidth]{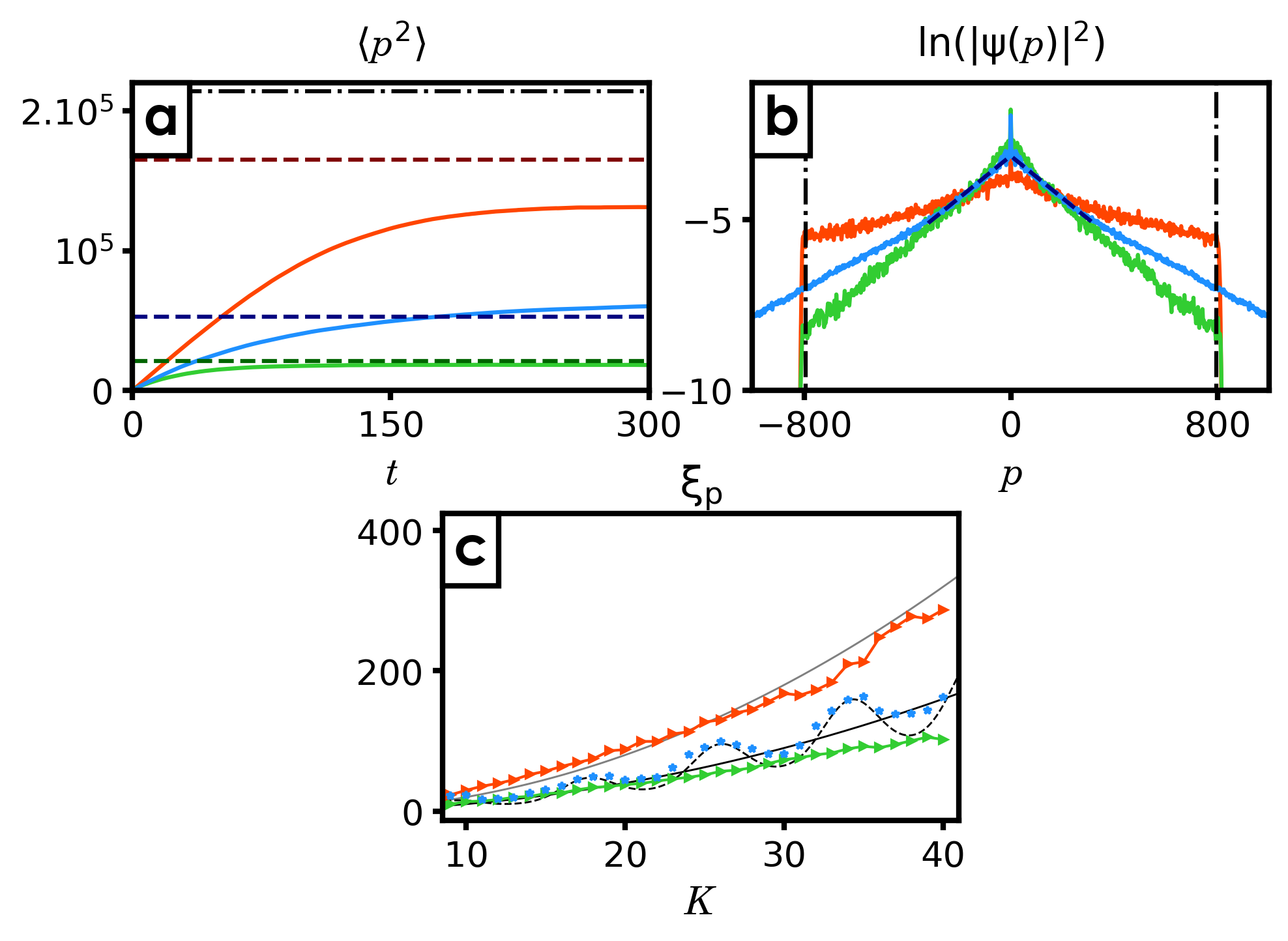}
	\caption{\textbf{Quantum dynamical localization in the kicked rotor and shaken rotor models.}  
\textbf{(a)} Variance \( \langle p^2 \rangle \) of a wave packet initially peaked at \( p = 0 \) as a function of time in the \( \varnothing \)-symmetric shaken rotor (red), the \( \mathsf{PT} \)-symmetric shaken rotor (green), and the kicked rotor (blue). The parameters used are \( K = 40 \), \( \hbar_\mathrm{eff} = 2.5 \), and \( N_H = 126 \), leading to \( L_p \approx 2\pi(2N_H+1) \approx 1589 \) for the shaken rotors.  
The black dash-dotted line represents the value \( \langle p^2 \rangle = L_p^2/12 \), corresponding to a uniform momentum distribution over the finite chaotic sea.  
For all three systems with these parameters, the saturation of \( \langle p^2 \rangle \) at large times is due to dynamical localization, with \( \langle p^2 \rangle \approx 2\xi_p^2 \), as indicated by the corresponding colored dashed lines. The values of \( \xi_p \) are extracted from a fit to the exponential decay of the probability density, as shown in \textbf{(b)}.  
\textbf{(b)} Long-time probability density in momentum space for the three systems, with parameters \( K = 40 \), \( \hbar_\mathrm{eff} = 2.5 \), and \( t = 512 \). The distributions exhibit an exponential decay, with fits shown as black dashed lines, yielding the localization length \( \xi_p \). The black dotted line represents the range of the chaotic sea of the shaken rotor models in momentum space, \( p \in [-L_p/2, L_p/2] \), where \( L_p = 2\pi(2N_H+1) \approx 1589 \).  
\textbf{(c)} Localization length \( \xi_p \) as a function of \( K \) for the three systems. The black dashed line represents the analytical prediction for the kicked rotor~\cite{SHEPELYANSKY1987103}. The black solid line corresponds to \( \frac{1}{4} \frac{K^2}{\hbar_\mathrm{eff}} \), while the gray solid line corresponds to \( \frac{1}{2} \frac{K^2}{\hbar_\mathrm{eff}} \).  
	}
	\label{fig:CompareKR_Quantum_Localization}
\end{figure}

Since the two shaken rotor models belong to different symmetry classes, their localization lengths are expected to differ~\cite{Pichard1990, PhysRevLett.69.217}. The \(\mathsf{PT}\)-symmetry-preserving shaken rotor belongs to the Orthogonal class, while the \(\varnothing\)-symmetry rotor belongs to the Unitary class. Denoting the localization lengths as \( \xi_p^O \) (Orthogonal) and \( \xi_p^U \) (Unitary), one expects the relation \( \xi_p^U \approx 2\xi_p^O \)~\cite{Pichard1990, PhysRevLett.69.217}. As a result, the saturation values of \( \langle p^2 \rangle \) differ between the two systems, as seen in Fig.~\ref{fig:CompareKR_Quantum_Localization} \textbf{a}.  

The localization length can be extracted by fitting the exponential decay of the momentum distribution after an evolution time exceeding the localization time, as shown in Fig.~\ref{fig:CompareKR_Quantum_Localization} \textbf{b}. As expected, the ratio of localization lengths in the two shaken rotor systems is approximately 2 (see Fig.~\ref{fig:CompareKR_Quantum_Localization} \textbf{c}).

\medskip

\subsubsection{CFS peak growth}

Finally, we describe the growth of the CFS peak in the shaken rotor, comparing it in particular to its well-known dynamics in the kicked rotor.  
As stated in the manuscript, a key interest of the shaken rotor model is that it allows the investigation of the CFS peak in both the classically bounded (\( L_p \ll \xi_p \)) and dynamically localized (\( L_p \gg \xi_p \)) regimes, as well as in different symmetry classes—here, the Orthogonal and Unitary classes.  

An important property of the CFS peak is that its growth is governed by the spectral form factor \( \mathcal{K}(t) \), see e.g.~\cite{Martinez_2023, Marinho_2018}. The form factor \( \mathcal{K}(t) \) is the Fourier transform of the two-point energy correlator, defined as  
\begin{equation}
    \mathcal{K}(t) = \frac{1}{N_S} \left\langle \sum_{n, m} e^{-i(\epsilon_n - \epsilon_m)t} \right\rangle = \frac{1}{N_S} \langle \lvert \text{tr}(\hat{U}^t) \rvert ^2 \rangle,
\end{equation}
where \( \hat{U} \) is the evolution operator, a unitary matrix of size \( N_S \times N_S \) with $N_S$ the numerical model system size, and \( \epsilon_n \) denotes its quasi-energies.  

The spectral form factor \( \mathcal{K}(t) \) is known analytically for certain Random Matrix Ensembles~\cite{LesHouches1989}. In the Gaussian Orthogonal Ensemble (GOE), it takes the form  
\begin{equation}
    \mathcal{K}^O(t) =
    \begin{cases}
        \frac{2t}{t_H} - \frac{t}{t_H} \log\left(1 + \frac{2t}{t_H}\right), & \text{if } t < t_H; \\
        \\
        2 - \frac{t}{t_H} \log\left(\frac{2t/t_H + 1}{2t/t_H - 1}\right), & \text{if } t \geq t_H,
    \end{cases}
\end{equation}
while in the Gaussian Unitary Ensemble (GUE), it is given by  
\begin{equation}
    \mathcal{K}^U(t) =
    \begin{cases}
        \frac{t}{t_H}, & \text{if } t < t_H; \\
        \\
        1, & \text{if } t \geq t_H,
    \end{cases}
\end{equation}
where \( t_H \) is the Heisenberg time.  

Conversely, for 1D Anderson localization, the form factor follows~\cite{Marinho_2018}  
\begin{equation}
    \mathcal{K}^O_{\text{loc}}(t) = \left( I_0 \left( \frac{2t_H}{t} \right) + I_1 \left( \frac{2t_H}{t} \right) \right) \exp \left( -\frac{2t_H}{t}\right),
\end{equation}
in the Orthogonal class, and  
\begin{equation}
    \mathcal{K}^U_{\text{loc}}(t) =  I_0 \left( \frac{t_H}{t}\right) \exp\left( - \frac{t_H}{t} \right),
\end{equation}
in the Unitary class, with $I_{0,1}$ the modified Bessel functions.  

\begin{figure}[h]
	\centering
	\includegraphics[width=\linewidth]{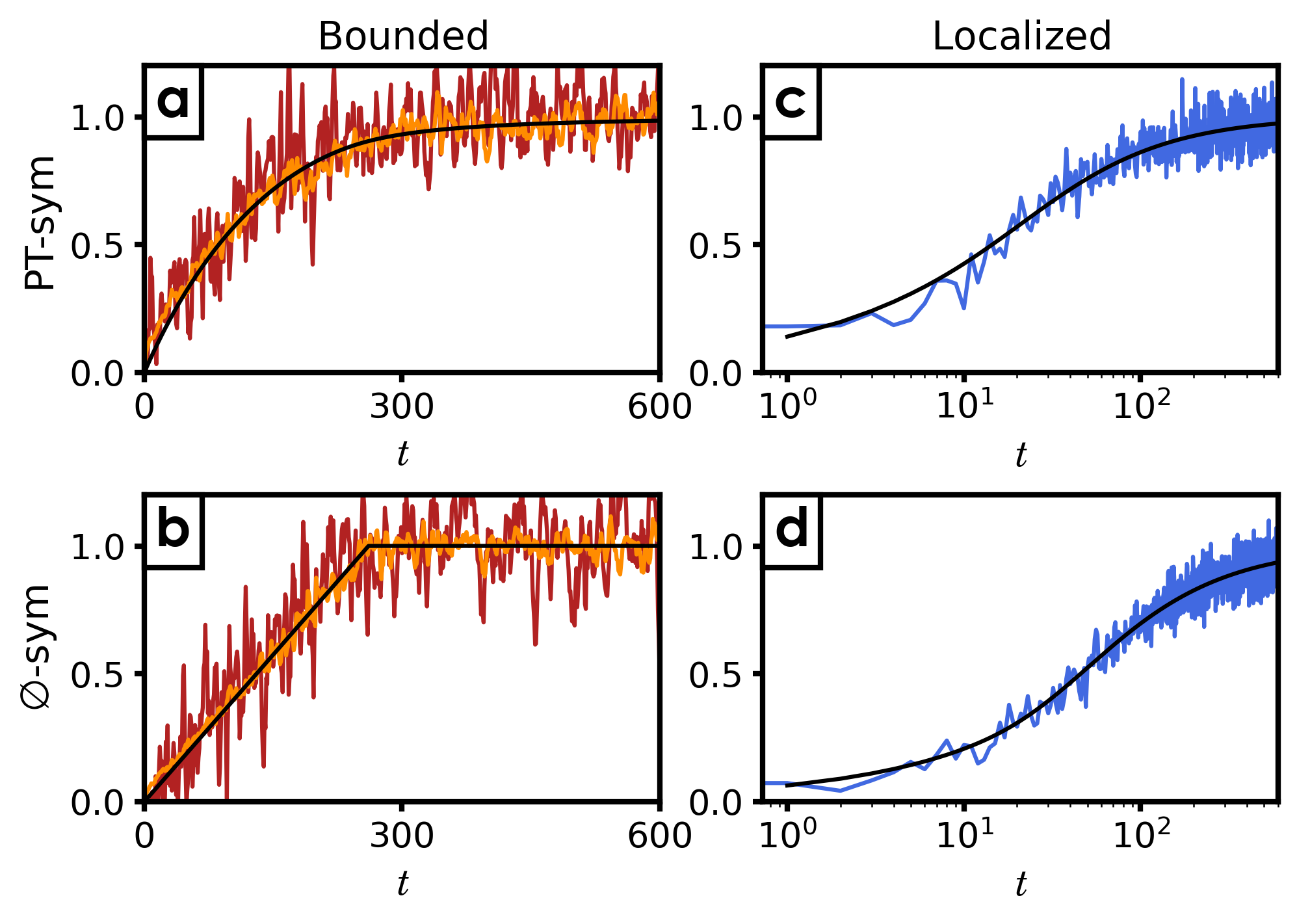}
	\caption{\textbf{Analytical and Numerical Form Factor and CFS contrast, in the $\mathsf{PT}$-sym and $\varnothing$-sym shaken rotor, in the localized and classically bounded regimes.} $\mathbf{(a-b)}$: In the classically bounded case: renormalized numerical CFS contrast $\mathcal C_{\mathrm{CFS}}(t)$ (brown curves), numerical computation of the Form Factor $\mathcal{K}(t)$ (orange curves), and analytical prediction of the Form Factor $\mathcal{K}^{O}(t)$ in $\mathbf{(a)}$ and  $\mathcal{K}^{U}(t)$  in $\mathbf{(b)}$ (black curves). Parameters: $K = 40$, $\hbar_\mathrm{eff} = 1$, $N_H = 17$, $L_p \approx 220$. We fit $\mathcal{K}(t)$ with the appropriate prediction to extract the Heisenberg time $t_H$ of the system: $\mathbf{(a)}$ $t_H \approx 257.7 \simeq L_p$, corresponding to the extension of the chaotic sea. $\mathbf{(b)}$ $t_H \approx 261 \simeq L_p$. $\mathbf{(c}$-$\mathbf{d)}$ Localized case: renormalized numerical CFS contrast $\mathcal C_{\mathrm{CFS}}(t)$ (blue curves) and analytical predictions $\mathcal{K}_\text{loc}^{O}(t)$ $\mathbf{(c)}$-$\mathcal{K}_\text{loc}^{U}(t)$ in $\mathbf{(d)}$ (black curves). Parameters: $K = 20$, $\hbar_\mathrm{eff} = 2.5$, $N_H = 128$, $L_p \approx 645$. We fit the renormalized numerical CFS contrasts with the analytical predictions to extract the time $t_H$. $\mathbf{(c)}$ Localized $\mathsf{PT}$-symmetric shaken rotor with $t_H \approx 16.18$, corresponding to $\xi_{p}/\hbar_\mathrm{eff}$. $\mathbf{(d)}$ Localized $\varnothing$-symmetric shaken rotor with $t_H \approx 40.05\simeq\xi_p/\hbar_\mathrm{eff}$.
	}
	\label{fig:MonteeCFS_KRF}
\end{figure}

The behavior of the CFS peak in the 1D Anderson localized regime was shown to correspond to \( \mathcal{K}^{O,U}_{\text{loc}}(t) \) in the kicked rotor \cite{PhysRevA.95.043626}. In Fig.~\ref{fig:MonteeCFS_KRF}, we show that this is also the case for the shaken rotor models in both the Orthogonal and Unitary symmetry classes, in the dynamically localized regime. Additionally, we consider the classically bounded case and demonstrate that it corresponds to \( \mathcal{K}^{O,U}(t) \) from random matrix theory. More precisely, we examine the time evolution of the CFS contrast. Starting from an initial state peaked at \( x_0 \), the normalized contrast \( \mathcal{C}_{CFS}(t) \) is defined as  
\begin{equation}
    \mathcal{C}_{CFS}(t) = (n(x_0,t) - 1)/(\lim_{t \to \infty} n(x_0,t) - 1),
\end{equation}
where \( n(x_0,t) \) is the averaged spatial probability density, given by  
\begin{equation}
    n(x_0,t) = N_S \overline{\lvert \psi (x_0,t) \rvert ^2},
\end{equation}
  The normalization coefficient \( \lim_{t \to \infty} n(x_0,t) - 1 \) is used in order to compare the CFS contrast with the form factor, which always reaches 1 at long times.  

We have computed the CFS contrast and the form factor in the two shaken rotor models and in both the classically bounded and localized regimes (see Fig.~\ref{fig:MonteeCFS_KRF}). In the classically bounded case, the growth of the CFS peak follows the form factor \( \mathcal{K}^{O,U}(t) \), and fitting the analytical prediction \( \mathcal{K}(t) \) yields a Heisenberg time \( t_H \) close to the chaotic sea size \( L_p \), as expected. In the localized case, we compute the CFS contrast on a logarithmic time scale. For the same system parameters, in the \(\mathsf{PT}\)-symmetric case, the characteristic time \( t_H \) is approximately half of that in the \(\varnothing\)-symmetric case, as expected. Moreover, these characteristic times match well with the localization lengths previously computed in Fig.~\ref{fig:CompareKR_Quantum_Localization}.

\subsection*{Average probabilities}
\label{sup:proba}

\begin{figure*}[t!]
	\centering
	\includegraphics[width=0.71\textwidth]{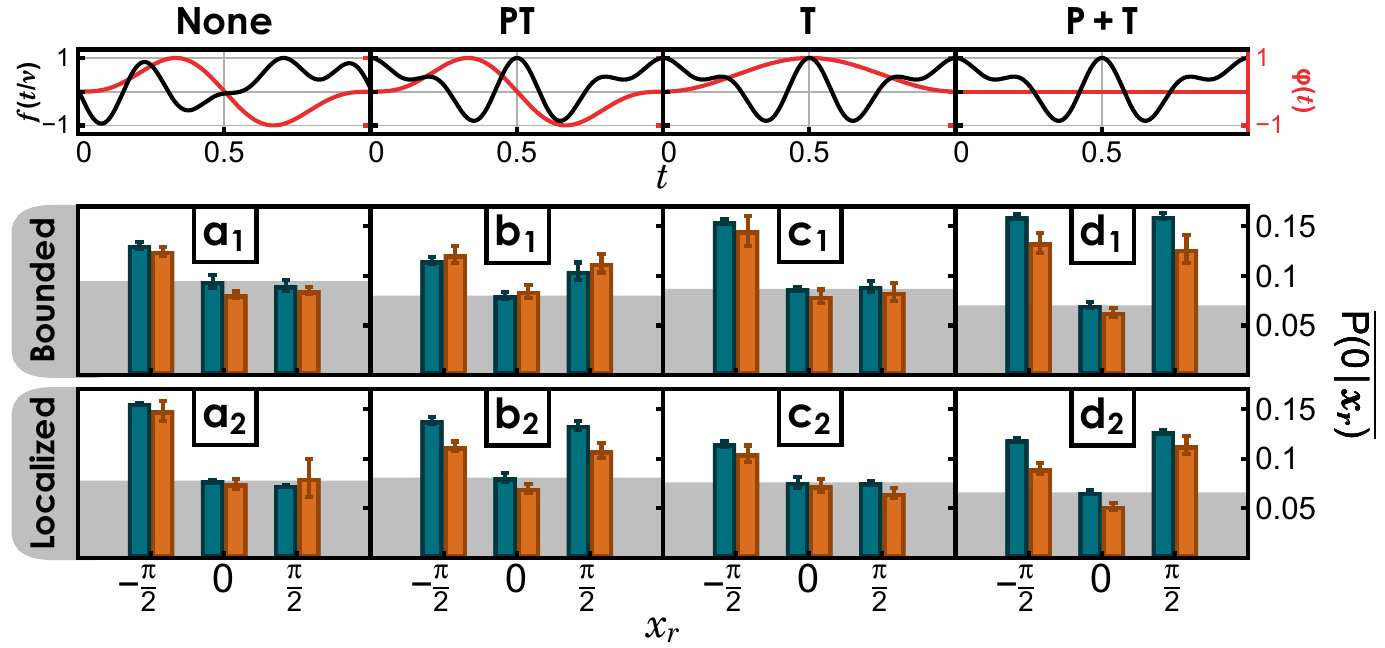}
	\caption{
    (top): Shape of periodic modulations of lattice depth (black) and position (red) corresponding to symmetries (none, $\mathsf{PT}$, $\mathsf{T}$, $\mathsf{P+T}$ for panels \textbf{(a-d)} respectively). (bottom) Experimental (orange) and numerical (blue) average probabilities $\overline{P(0,x_r)}$ obtained for different symmetries of the modulation, and in two different localization regimes:  bounded $\xi_{\rm loc}/L\gtrsim 1$, and localized $\xi_{\rm loc}/L\lesssim 0.1$, where $\xi_{\rm loc}=\xi_p/\hbar_\mathrm{eff}$ is the localization length and $L=L_p/\hbar_\mathrm{eff}$ the extension of the chaotic sea, in units of $\hbar_\mathrm{eff}$. The signals are obtained using the method presented in Figure 2 of the main text with an additional average over 3 modulations periods. Parameters are the same as in Figure 3 of the main text.
    }
	\label{fig:Figure_3_v7_proba}
\end{figure*}

On Figure~\ref{fig:Figure_3_v7_proba}, we show the average probability values at $\ell=0$, and for $x_r=0,\pm\pi/2$, that were obtained experimentally (orange bars) and from the full numerical modeling (blue bars). It is from these distributions that we compute the CBS and CFS contrasts, and the regimes and parameters are the ones of Figure 3 of the main article.

\subsection*{CBS dynamics}
\label{sup:CBS}

\begin{figure}[t!]
	\centering
	\includegraphics[width=0.9\linewidth]{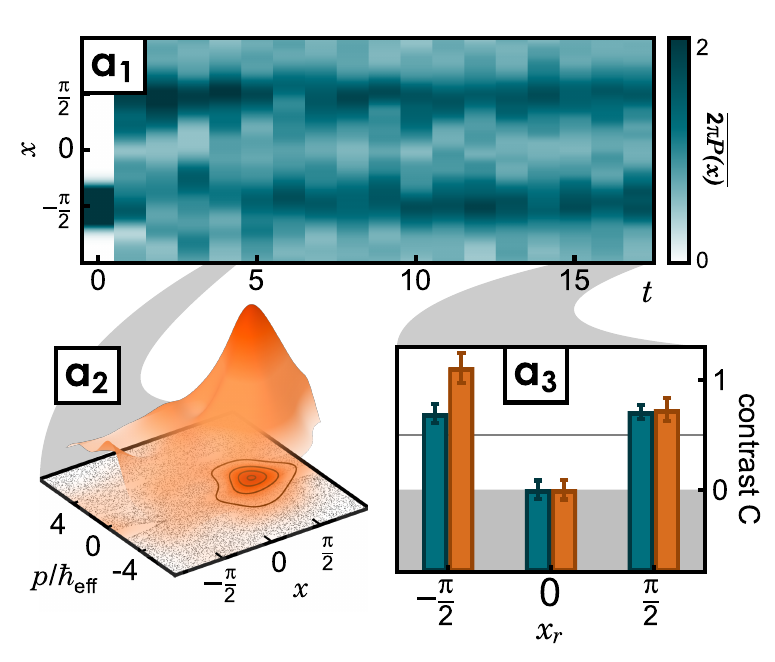}
	\caption{\textbf{CBS dynamics in $\mathsf{PT}$ symmetry.}  $\mathbf{(a_1)}$ Evolution of the  averaged spatial density.  $\mathbf{(a_2)}$ Experimentally reconstructed average Husimi distribution at a short time $N=4$ (with contour plot projections lines at $50, 75, 90$ and $99\%$ of maximum) showing the presence of CBS.$\mathbf{(a_3)}$ Phase space rotation signals obtained at a longer time ($N=[15,16,17]$), where both CBS and CFS are present. Parameters: $\mathsf{PT}$ symmetry, $s=19.7\pm0.21$, $\nu=30.888\,$kHz, $N_H=5$, $A_\varphi=1$, $K=8.76\pm0.74$, $\hbar_{\rm eff}=3.3$, $L=25.06\pm1.94$, $\xi_{\mathrm{loc}}=2.45\pm1.21$ .
	}
	\label{fig:Fig6_MonteeCBS}
\end{figure}

The CBS peak appears on short timescales, corresponding to the elastic scattering time $t_{scatt}\sim1$, in contrast to the CFS peak, which arises around the Heisenberg time, $t_H$, determined by localization \cite{PhysRevA.95.043626}. 
We can compare their dynamics in $\mathsf{PT}$-symmetry (see Fig. \ref{fig:Fig6_MonteeCBS}), where both peaks are expected.
The figure illustrates, for a localized regime, an experimentally reconstructed Husimi distribution at short time $N=4$, showing a single peak at $x=\pi/2$ corresponding to the CBS. At longer times ($N=[15,16,17]$), using the phase space rotation method, both CBS and CFS are observed.

\end{appendix}

\end{document}